\documentclass[journal,draftclsnofoot,onecolumn,12pt]{IEEEtran}

\usepackage{acronym}
\usepackage{amsfonts}
\usepackage[dvips]{graphicx}
\usepackage{times}
\usepackage{cite}
\usepackage{amsmath}
\usepackage{array}
\usepackage{amssymb}
\usepackage{stfloats}
\usepackage{diagbox}
\usepackage{graphicx}
\usepackage{footnote}
\usepackage{amsthm}
\usepackage{booktabs}
\usepackage{array}
\usepackage[ruled,vlined]{algorithm2e}
\usepackage{subeqnarray}
\usepackage{cases}
\usepackage{threeparttable}
\usepackage{color}
\usepackage{epstopdf}
\usepackage{multirow}
\usepackage{tabularx}
\usepackage{enumerate}
\usepackage{multicol}
\usepackage{hyperref}
\usepackage{subfig}
\usepackage{caption}

\def\bb0{{\mathbb{0}}}

\def\ba{{\mathbf{a}}}
\def\bb{{\mathbf{b}}}
\def\bc{{\mathbf{c}}}

\def\bh{{\mathbf{h}}}

\def\bm{{\mathbf{m}}}
\def\bn{{\mathbf{n}}}

\def\br{{\mathbf{r}}}

\def\bv{{\mathbf{v}}}
\def\bw{{\mathbf{w}}}
\def\bx{{\mathbf{x}}}
\def\by{{\mathbf{y}}}
\def\bz{{\mathbf{z}}}
\def\b0{{\mathbf{0}}}

\def\bA{{\mathbf{A}}}

\def\bF{{\mathbf{F}}}

\def\bI{{\mathbf{I}}}

\def\bW{{\mathbf{W}}}

\def\bbC{{\mathbb{C}}}

\def\cN{\mathcal{N}}

\def\cQ{\mathcal{Q}}
\def\cR{\mathcal{R}}

\def\sfP{\mathsf{P}}

\def\sf0{{\mathsf{0}}}

\def\rmD{\mathrm{D}}

\def\sign{{\mathrm{sign}}}

\def\rm0{{\mathrm{0}}}

\acrodef{CSI}[CSI]{channel state information}
\acrodef{CSIT}[CSIT]{channel state information at the transmitter}
\acrodef{CSIR}[CSIR]{channel state information at the receiver}
\acrodef{MIMO}[MIMO]{multiple-input multiple-output}
\acrodef{SISO}[SISO]{single-input single-output}
\acrodef{MISO}[MISO]{multiple-input single-output}
\acrodef{SIMO}[SIMO]{single-input multiple-output}
\acrodef{ADCs}[ADCs]{analog-to-digital convertors}
\acrodef{SNR}[SNR]{signal-to-noise ratio}
\acrodef{AWGN}[AWGN]{additive white Gaussian noise}
\acrodef{MRT}[MRT]{maximal ratio transmission}
\acrodef{DFT}[DFT]{Discrete Fourier Transform}
\acrodef{ULA}[ULA]{uniform linear array}
\acrodef{UPA}[UPA]{uniform planar array}
\acrodef{LS}[LS]{least squares}
\acrodef{ALMMSE}[ALMMSE]{approximate linear minimum mean squared error}
\acrodef{QIHT}[QIHT]{quantized iterative hard thresholding}
\acrodef{QIST}[QIST]{quantized iterative soft thresholding}
\acrodef{SVD}[SVD]{singular value decomposition}

\newcommand{\blc}[1]{\color{blue}#1}

\usepackage{bm}
\begin{document}

\title{Beamspace Channel Estimation for Wideband Millimeter-Wave MIMO: A Model-Driven Unsupervised Learning Approach}

\author{Hengtao He,
Rui Wang, Weijie Jin,
Shi Jin,~\IEEEmembership{Senior Member,~IEEE,}
Chao-Kai Wen,~\IEEEmembership{Senior Member,~IEEE,}
and Geoffrey Ye Li,~\IEEEmembership{Fellow,~IEEE}
\thanks{H.~He, R. Wang, W. Jin and S.~Jin are with the National Mobile Communications Research
Laboratory, Southeast University, Nanjing 210096, China (e-mail: hehengtao@seu.edu.cn, wang\_rui@seu.edu.cn, jinweijie@seu.edu.cn, and jinshi@seu.edu.cn).}
\thanks{C.-K.~Wen is with the Institute of Communications Engineering, National
Sun Yat-sen University, Kaohsiung 804, Taiwan (e-mail: chaokai.wen@mail.nsysu.edu.tw).}
\thanks{G. Y. Li is with the Department of Electrical and Electronic Engineering, Imperial College London. (email: geoffrey.li@imperial.ac.uk).
}
}
\maketitle

\begin{abstract}
Millimeter-wave (mmWave) communications have been one of the promising technologies for future wireless networks that integrate a wide range of data-demanding applications. To compensate for the large channel attenuation in mmWave band and avoid high hardware cost, a lens-based beamspace massive multiple-input multiple-output (MIMO) system is considered. 
However, the beam squint effect  in wideband mmWave systems makes channel estimation very challenging, especially when the receiver is equipped with a limited number of radio-frequency (RF) chains.  Furthermore, the real channel data cannot be obtained before the mmWave system is used in a new environment, which makes it impossible to train a deep learning (DL)-based channel estimator using real data set beforehand. To solve the problem, we propose a model-driven unsupervised learning network, named learned denoising-based generalized expectation consistent (LDGEC) signal recovery network.  By utilizing the Stein's unbiased risk estimator loss, the LDGEC network can be trained only with limited measurements corresponding to the pilot symbols, instead of the real channel data. Even if designed for unsupervised learning, the LDGEC network can be supervisingly trained with the real channel via the denoiser-by-denoiser way. The numerical results demonstrate that the LDGEC-based channel estimator significantly outperforms state-of-the-art compressive sensing-based algorithms when the receiver is equipped with a small number of RF chains and low-resolution ADCs.
\end{abstract}


\IEEEpeerreviewmaketitle

\section{Introduction}
MmWave communications have been considered as a promising technology to support the very high data rate in
future wireless communications since it can provide a tenfold increase in the bandwidth \cite{swindlehurst2014millimeter,bai2015coverage,RHeath2016JSTSP}.
However, as the carrier frequency increases, the mmWave signals suffer from much more severe attenuation, which becomes a vital issue in mmWave communications.  Leveraging the large antenna arrays employed at the transmitter and receiver, massive multiple-input multiple-output (MIMO) can perform directional beamforming to achieve a high beamforming gain, which helps overcome large pathloss of mmWave signals and guarantees sufficient received signal-to-noise ratio (SNR). However, the hardware cost and power consumption both increase with {\blc the} number of RF chains, which is sometimes unaffordable if a dedicated RF chain is used for each of a huge number of antennas. To reduce the number of required RF chains, we can resort to beamspace massive MIMO with a discrete lens array (DLA), which has been first proposed in \cite{beamspace2013millimeter} and successfully employed in millimeter-wave (mmWave) communications. {\blc However}, the number of RF chains is much smaller than that of antennas, and we cannot directly observe the complete channel in the baseband \cite{alkhateeb2014channel}, thus incurring challenges for  beamspace channel estimation.
\subsection{Related Work}
For beamspace channel estimation, several works utilize compressive sensing (CS) techniques \cite{alkhateeb2014channel,Beam2016Sayeed, estimation2016Yang, SD, SCAMPI} in mmWave band.
The training-based scheme in \cite{Beam2016Sayeed} first scans all the beams and retains only a few strong ones. Then, the least-square (LS) algorithm is employed for estimating the reduced-dimensional beamspace channel. In \cite{estimation2016Yang}, a modified version of \cite{Beam2016Sayeed} reduces the overhead of beam training by simultaneously scanning several beams with the help of power splitters at the BS. However, the aforementioned algorithms are not optimized for lens-based mmWave systems because the lens antenna array has energy-focusing capability, and the received signal matrix from the lens antenna array is characterized by sparsity and concentration. The support detection based scheme in \cite{SD} further reduces the pilot overhead, which directly estimates the channel support by exploiting the sparsity of the beamspace channel. In \cite{SCAMPI}, the channel matrix is regarded as a 2-dimensional (2D) natural image and is then estimated by the cosparse analysis approximate message passing (SCAMPI) algorithm derived from the image recovery field. The SCAMPI algorithm models the channel as a sparse generic $L$-term Gaussian mixture (GM) probability distribution and uses the expectation-maximization (EM) algorithm to learn the GM parameters from the current estimated data. 

{\blc Previous works only address narrowband mmWave systems. For wideband mmWave massive MIMO systems, the physical propagation delays of electromagnetic waves traveling across the whole array will become large and comparable to the time-domain sample period. In such a case, different antenna elements will receive different time-domain symbols, which is known as the spatial-wideband effect \cite{beam2018mag} and causes beam squint in the frequency domain. As a result, the AoAs (AoDs) will become frequency-dependent; thereby,  channel estimation becomes very challenging, especially in mmWave band. The successive support detection (SSD) technique  proposed in \cite{beam2019Gao} applies successive interference cancelation to estimate the channel. The main idea is that each sparse path component has frequency-dependent support determined by its spatial direction and is estimated using beamspace windows. However, some important characteristics of mmWave channels, such as sparsity and channel correlation between adjacent antennas and subcarriers, are not considered. These characteristics are significant for performance improvement in channel estimation, but are difficult to be characterized by traditional model-based method.

Recently, deep learning (DL) has been applied to physical layer communications \cite{DL2018Qin,Modeldriven18DL,CSINet,OAMP-Net2,DL2018HE,DL2020Qi,JSTSP2019Dong, DL2020precoding}, such as channel state information (CSI) feedback \cite{CSINet}, signal detection \cite{DL2OFDM,OAMP-Net2}, channel estimation \cite{DL2018HE,DL2020Qi,JSTSP2019Dong}, precoder design \cite{DL2020precoding}, {\blc traffic analysis \cite{Mobile2019,Rago2020,Aceto2019},} and end-to-end transceiver design \cite{DL2018Hoydis,GAN}. DL-based physical layer communications may be data-driven and model-driven \cite{Modeldriven18DL}. By incorporating domain knowledge into network design, model-driven DL can reduce the demand for computing resources
and training time, which is more attractive for wireless communications \cite{DL2018HE}. As a promising model-driven DL approach, deep unfolding has been first proposed  in \cite{Deep14unfolding} and  applied to sparse signal recovery \cite{TISTA} and image processing \cite{Deep19Eldar}. The main idea is unfolding the iterative algorithm into a deep neural network and adding learnable parameters, which has been successfully applied to physical layer communications \cite{Modeldriven18DL,DL2018HE,Deep19studer}. For example, the deep unfolding-based channel estimator has been successfully applied for narrowband beamspace mmWave massive MIMO systems in \cite{DL2018HE}.  It outperforms  state-of-the-art CS-based algorithms and can achieve excellent performance even with a small number of RF chains. However, the existing DL-based channel estimator utilizes a supervised way \cite{DL2018HE,DL2020Qi,JSTSP2019Dong}, thereby a large number of real channel data are required to train the network, which defeats the point of channel estimation in the first place. This is because we cannot obtain the true channel data for training the network when the system is equipped in a practical environment. Therefore, how to train the DL-based channel estimator without the true channel data is significantly important. }

\subsection{Contributions}

In this study, we develop a DL-based channel estimator for lens-based mmWave massive MIMO systems. Instead of considering supervised learning for narrowband beamspace channel estimation \cite{DL2018HE}, we investigate unsupervised learning for a wideband system and take the beam squint effect into consideration. To the best of our knowledge, this paper is the first study applies model-driven unsupervised DL network into wideband mmWave beamspace massive MIMO systems and considers the beam squint effect. {\blc The main contributions  are summarized as follows.}

\begin{itemize}

  \item
  {\blc We first formulate the wideband beamspace channel estimation problem as a compressed image recovery problem.}
  By incorporating an advanced denoising convolutional neural network (DnCNN) into the generalized expectation consistent signal recovery (GEC-SR) algorithm \cite{JSTSP18He,TSP20Wang}, we develop a model-driven DL network, named the {\blc learned} denoising-based GEC (LDGEC) network. The LDGEC network uses the Steins unbiased risk estimator (SURE) as the loss function; thereby, it can be trained only with the received signals not the real channel data.  By utilizing \emph {layer-by-layer training}, the LDGEC-based estimator can significantly outperform state-of-the-art CS algorithms even without the real channel data.

  \item
  Even if designed for unsupervised learning, the LDGEC network can also be supervisingly trained with the real channel data, {\blc thereby further improve channel estimation performance with available channel data}. In this case, we can train the DnCNN denoiser in the \emph{denoiser-by-denoiser} way, where the DnCNN denoiser is trained independently without including the whole GEC algorithm, thereby reducing the training complexity significantly.

  \item To further reduce the cost and power consumption, we investigate the LDGEC-based channel
  estimator for systems with hardware-friendly low-resolution ADCs. {\blc Numerical results demonstrate that little performance loss is caused for LDGEC-based channel estimator }when the mmWave beamspace system is with low-resolution ADCs and reduced number of RF chains.

\end{itemize}

\emph{Notations}---For any matrix $\mathbf{A}$, $\mathbf{A}^{T}$ and ${ \mathrm{tr}}(\mathbf{A})$ denote  the transpose and the trace of $\mathbf{A}$, respectively. In addition, $\mathrm{Diag}(\mathbf{v})$ is the diagonal matrix with  $\mathbf{v}$ on the diagonal, and $\mathbf{d}(\mathbf{Q})$ is the diagonalization operator, which returns a constant vector containing the average diagonal elements of $\mathbf{Q}$. Furthermore, $\mathbb{E}\{\cdot\}$ represents the expectation operator. A circular complex Gaussian with mean $\boldsymbol{\mu}$ and covariance $\boldsymbol{\Omega}$ can be described by the probability density function:
\begin{equation*}
  \mathcal{N}_{\mathbb{C}}(\mathbf{z};\boldsymbol{\mu},\boldsymbol{\Omega})=\frac{1}{\mathrm{det}(\pi \boldsymbol{\Omega})}
  e^{-(\mathbf{z}-\boldsymbol{\mu})^{H}\boldsymbol{\Omega}^{-1}(\mathbf{z}-\boldsymbol{\mu})}.
\end{equation*}
We use $\rmD z$ to denote the real Gaussian integration measure
\begin{equation*}
  Dz=\phi(z)dz, \quad \phi(z)\triangleq\frac{1}{\sqrt{2\pi}}e^{-\frac{z^{2}}{2}},
\end{equation*}
$\rmD z_{c}=\frac{e^{-|z|^{2}}}{\pi}dz$ to denote the complex Gaussian integration measure, $\Phi(x)\triangleq \int_{-\infty}^{x} \rmD z$
to denote the cumulative Gaussian distribution function.

The remaining part of this paper is organized as follows. Section \ref{system_model} formulates the wideband beamspace channel estimation as a compressed image recovery problem. Next, a model-driven DL network is provided for beamspace channel estimation using SURE loss in Section \ref{without_data}. Furthermore, the network can also be trained with the real channel data in Section \ref{with_data}. Then, numerical results are presented in Section \ref{simulation}. Finally, Section \ref{con} concludes the paper.
\begin{figure*}[t]
  \centering
  \includegraphics[width=15cm]{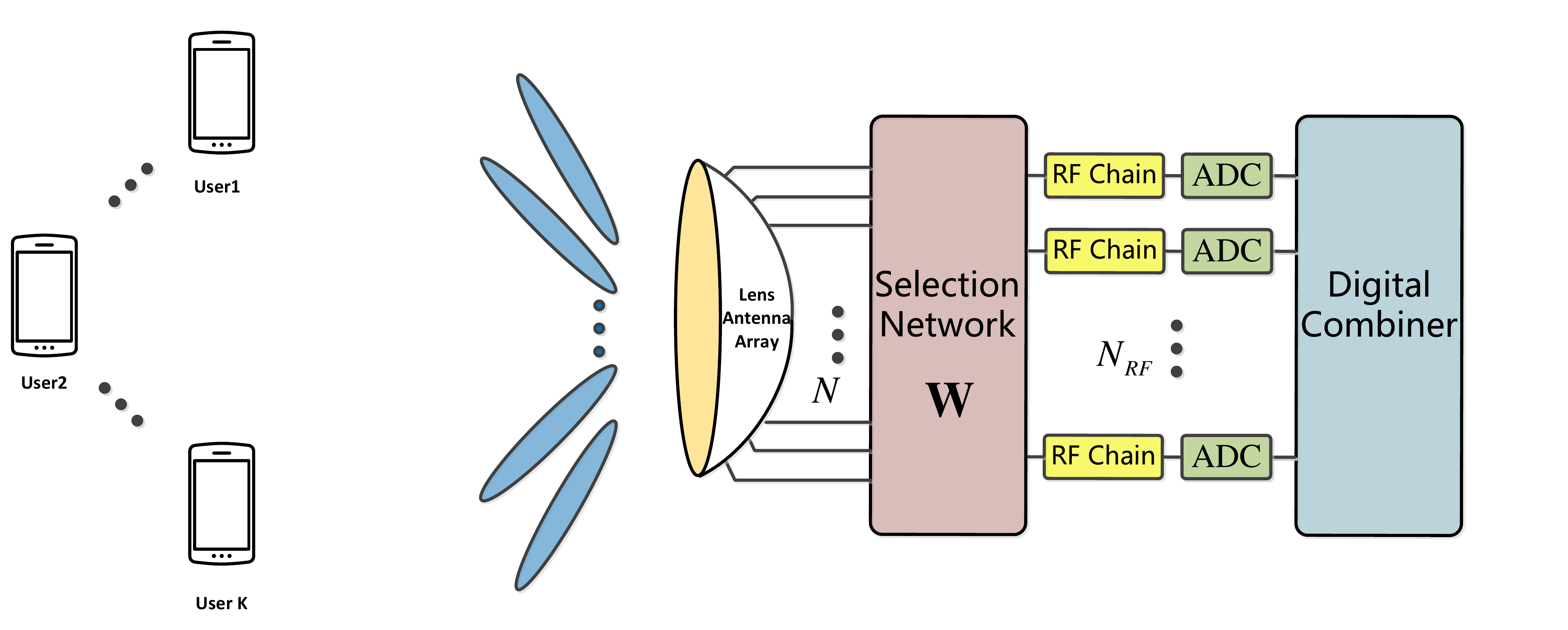}
  \caption{.~~The architecture of lens-based wideband beamspace mmWave MIMO-OFDM system.}\label{fig:system_model}
\end{figure*}
\section{System Model and Problem Formulation}\label{system_model}
In this section, we first present the lens-based mmWave MIMO-OFDM systems. After introducing the beam squint effect, we formulate the wideband beamspace channel estimation as a compressed image recovery problem.
\subsection{Beamspace channel model}\label{sub:channel}

As illustrated in Fig.1, we consider an uplink wideband bemspace mmWave MIMO-OFDM system, where the BS employs an $N$-element lens antenna array and $N_{RF}$ RF chains to simultaneously serve $K$ single-antenna users. Applying the classical Saleh-Valenzuela channel model \cite{SV2019Saleh}, the spatial channel $\bh_{m} \in \bbC^{N\times1}$ at sub-carrier $m$ is given by
\begin{equation}\label{eq:h}
  \bh_{m} = \sqrt{\frac{N}{L}}\sum_{l=1}^{L}\alpha_{l}e^{-j2\pi\tau_{l}f_{m}}\ba(\phi_{l,m}),
\end{equation}
for $m=1,2,\ldots, M$ where $L$ is the number of resolvable paths, $\alpha_{l}$ and $\tau_{l}$ are the complex gain and the time delay of the $l$-th path, respectively. Furthermore, $\ba(\phi_{l,m})$ is the array response vector and $\phi_{l,m}$ is the spatial direction at sub-carrier $m$ defined as
\begin{equation}\label{eqn:phi}
\phi_{l,m} = \frac{f_{m}}{c}d \sin\theta_{l},
\end{equation}
where $f_{m} = f_{c} + \frac{f_{s}}{m}(m-1-\frac{M-1}{2})$ is the frequency of sub-carrier $m$ with $f_{c}$ and $f_{s}$ representing the carrier frequency and bandwidth, respectively. Furthermore, $c$ is the speed of light, $\theta_{l}$ is the physical direction, and $d$ is the antenna spacing, which is usually designed according to the carrier frequency as $ d =c/2f_{c}$.  Consider a uniform linear lens array in the BS, the array response vector $\ba(\phi_{l,m})$ is given by,
\begin{align}\label{eqULA}
  \ba(\phi_{l,m}) & = \frac{1}{\sqrt{N}}[e^{-j2\pi\phi_{l,m}(-\frac{N-1}{2})}, e^{-j2\pi\phi_{l,m}(-\frac{N+1}{2})}, \ldots, \\  \nonumber
   &  e^{-j2\pi\phi_{l,m}(\frac{N-1}{2})}]^{T}.
\end{align}

The conventional channel in the spatial domain in (\ref{eq:h}) can be transformed to the beamspace domain by employing a carefully designed lens antenna array, as shown in Fig. \ref{fig:system_model}. Specifically,  this lens antenna array plays the role of an $N$-element spatial discrete Fourier transform (DFT) matrix
$\bF$, which contains the array response vectors of $N$ orthogonal directions (beams) covering the entire space as
\begin{equation}\label{eq:DFT}
  \bF = [\ba(\bar{\phi}_{1}), \ba(\bar{\phi}_{2}),\ldots ,\ba(\bar{\phi}_{N})],
\end{equation}
where $\bar{\phi}_{n}=\frac{1}{N}(n-\frac{N+1}{2})$ for $n=1,2,\ldots,N$ are the spatial directions pre-defined by the lens antenna array. Accordingly, the wideband beamspace channel $\tilde{\bh}_{m}$ at sub-carrier $m$ can be expressed as

\begin{equation}\label{eq:beam}
  \tilde{\bh}_{m}  = \bF^{H} \bh_{m} = \sqrt{\frac{N}{L} } \sum_{l=1}^{L}\alpha_{l}e^{-j2\pi\tau_{l}f_{m}}\tilde{\bc}_{l,m},
\end{equation}
where $\tilde{\bc}_{l,m}$ denotes the $l$-th path component at sub-carrier $m$ in the beamspace, and $\tilde{\bc}_{l,m}$ is determined by $\phi_{l,m}$ as
\begin{align}\label{eq:c}
  \tilde{\bc}_{l,m} & = \bF^{H}\ba(\phi_{l,m}) \\  \nonumber
   & = [\Xi(\phi_{l,m}-\bar{\phi}_{1}), \Xi(\phi_{l,m}-\bar{\phi}_{2}),\ldots,\Xi(\phi_{l,m}-\bar{\phi}_{N})]^{T},
\end{align}
where $\Xi(x) = \frac{\sin N 2 \pi x}{\sin \pi x}$ is the Dirichlet sinc function.

\subsection{Beam Squint}\label{sub:beam_squint}
Before formulating the wideband beamspace channel estimation problem, we introduce the beam squint effect \cite{beam2018mag}. Based on the power-focusing capability of $\Xi(x)$, we know that most of the power of $\tilde{\bc}_{l,m}$ is focused on only small number of elements. Additionally, due to the limited scattering in the mmWave systems, the number of resolvable paths, $L$, is generally small. However, the beam power distribution of the $l$-th path
component will be different at different sub-carriers, i.e., $\tilde{\bc}_{l,m_{1}} \neq \tilde{\bc}_{l,m_{2}} $
for $m_{1} \neq m_{2}$, since $\phi_{l,m}$ is frequency-dependent in wideband mmWave systems (i.e., $f_{m}\neq f_{c}$).
This effect is termed as beam squint \cite{beam2018mag}, which is a key difference between wideband and narrowband systems.

\begin{figure*}[t]
\begin{minipage}{3in}
  \centerline{\includegraphics[width=3.1in]{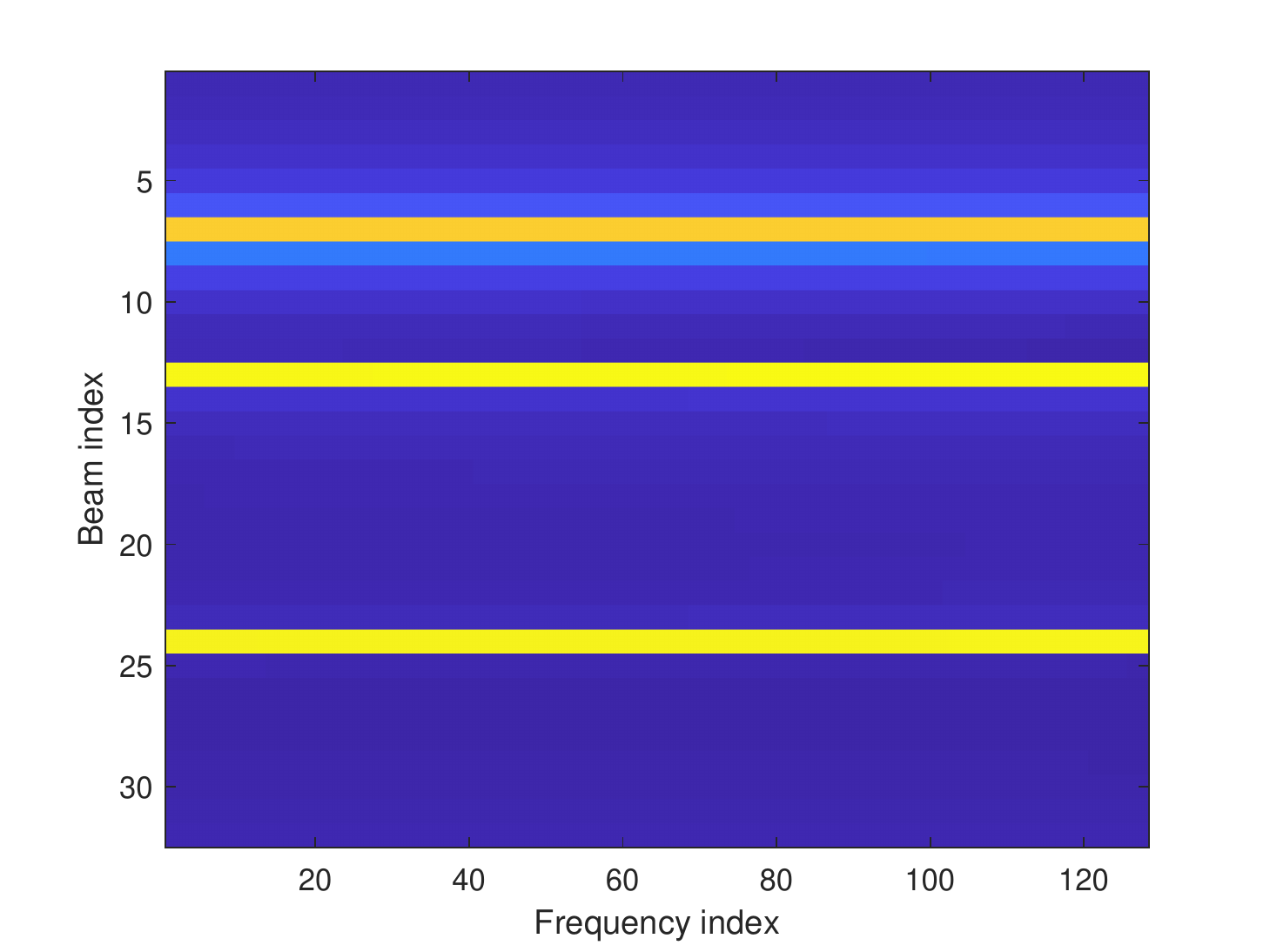}}
  \centerline{(a) Narrowband system}\label{fig:channela}
\end{minipage}
\hfill
\begin{minipage}{3in}
  \centerline{\includegraphics[width=3.1in]{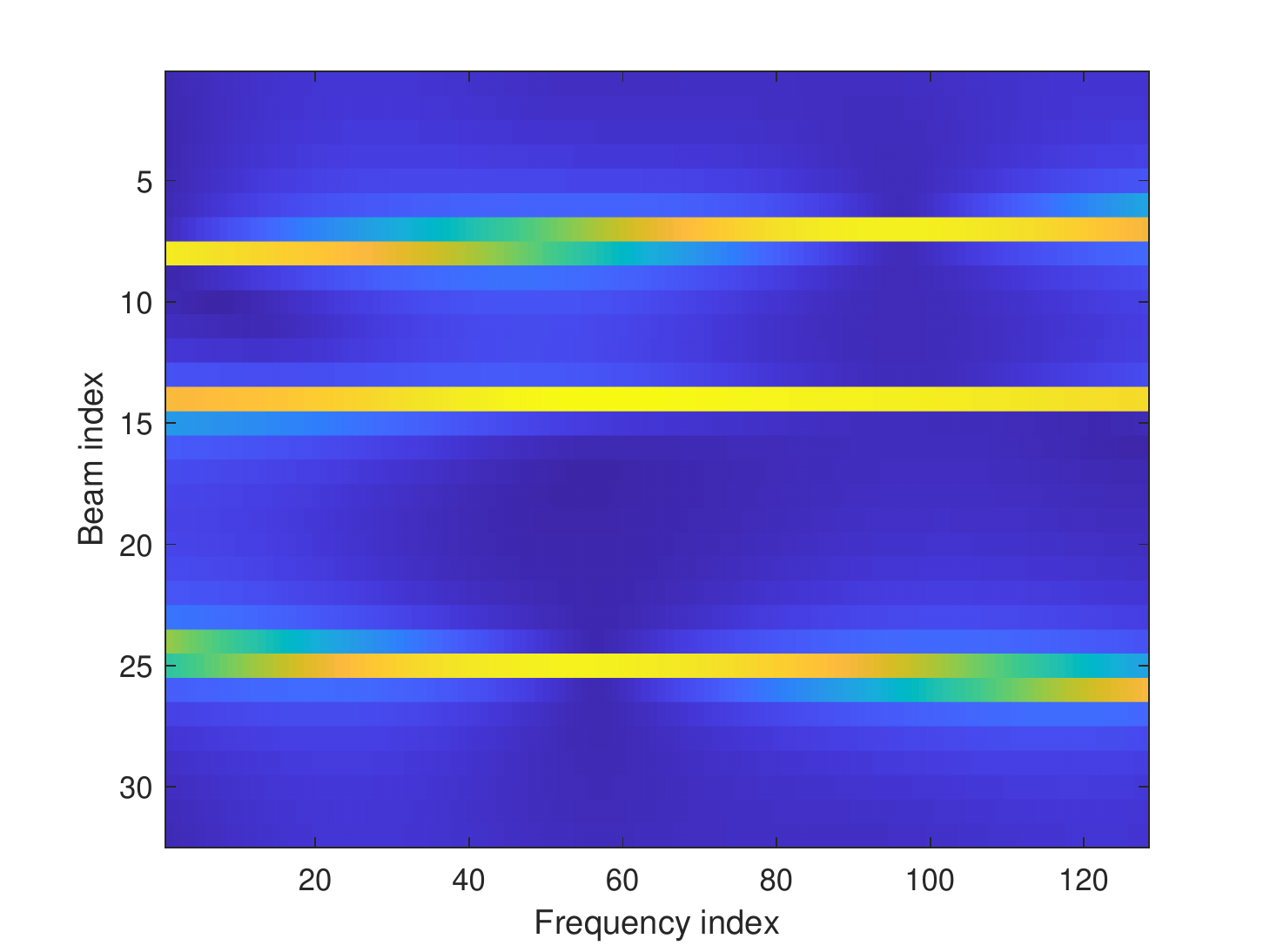}}
  \centerline{(b) Wideband system} \label{fig:channelb}
\end{minipage}
\caption{.~~An illustration of a beam-frequency channel, where $L=3$, $M=128$, $N=32$, $f_{c}=28$ GHz, $f_{s}=4$ GHz.}
\label{fig:channel}
\end{figure*}
{\blc
To show the beam squint effect, we present the energy diagram of the beam-frequency channel matrix in narrowband and wideband systems. We consider a beamspace system with $L=3$, $M=128$, $N=32$, $f_{c}=28$ GHz. Furthermore, we set $f_{s}=4$ GHz in a wideband system and  $f_{s}=20$ MHz in the narrowband system. As illustrated in Fig.\,\ref{fig:channel}, the beam power distribution of the $l$-th path component is almost similar at different sub-carriers in the narrowband system. Therefore, the beamspace channel support (the index of non-zero elements) at different frequencies can be assumed to be the same.
}

Owing to the beam squint effect, the beam power distribution in wideband systems varies significantly over frequency. {\blc
We denote the beam channel vector corresponding to the $l$-th path and the $m$-th frequency  as $\tilde{\mathbf{h}}_{l,m}$. As shown is in Fig.\,\ref{fig:channel}, the index of the strongest element in $\tilde{\mathbf{h}}_{3,0}$ (i.e., the yellow bar at the bottom of the Fig.\,\ref{fig:channel}) is $24$, while the index of the strongest element in $\tilde{\mathbf{h}}_{3,256}$ is $26$. Thus, the beamspace channel supports at different frequencies are different. The characteristic of the beam-frequency matrix will bring a significant challenge for wideband beamspace channel estimation. Furthermore, the channel correlation between adjacent antennas and subcarriers is subtle, which is difficult to be characterized by the traditional approaches. Conversely, DL has the powerful capability to learn the correlation from the data, which is more promising for wideband channel estimation involved in the beam squint effect.
}

\subsection{Problem Formulation}
In uplink channel estimation, the user devices transmit pilot sequences to the BS, and the
channel is assumed to remain unchanged during this period. We use the orthogonal pilot sequence for different users. Therefore, the channel estimation can be performed for each user independently. Considering a specific user, the  pilot at sub-carrier $m$ in instant $q$, $s_{m,q}$, is transmitted. The received signal vector $\by_{m,q} \in \mathbb{C}^{N \times 1}$ at the BS is given by
\begin{equation}\label{eq:y}
  \by_{m,q}=\bW_{q}\tilde{\bh}_{m}s_{m,q}+\bW_{q}\bn_{m,q},
\end{equation}
for $m = 1,2,\ldots,M$ and $q = 1,2,\ldots,Q$, where $\bn_{m,q}\sim \mathcal{N}_{\bbC}(\mathbf{0},\sigma_n^2\bI)$ represents a Gaussian noise vector. $\bW_{q} \in \bbC^{N_{RF}\times N}$ is the adaptive selection network but fixed for different sub-carriers. We set $s_{m,q}=1$ for convenience as the pilot signal is known at the receiver side. Thus, the received signal $\bar{\by}_{m}$ in $Q$ instants is given by
\begin{equation}\label{eq:y_all}
  \bar{\by}_{m}=[\by_{m,1}^{T}, \ldots, \by_{m,Q}^{T}]^{T}=\bar{\bW}\tilde{\bh}_{m}+\bW\bn_{m},
\end{equation}
where $\bar{\bW} = [\bW_{1}^{T}, \bW_{2}^{T},\ldots,\bW_{Q}^{T}]^{T} \in \bbC^{QN_{RF}\times N} $ and $\bn_{m}^{\mathrm{eq}}=[\bn_{m,1}^{T}, \ldots, \bn_{n,Q}^{T}]^{T}$. In this paper, low-cost $one$-bit phase shifters are utilized in the adaptive selection network $\bW_{q}$. Therefore, the elements of $\bar{\bW}$ are randomly selected from the set $\frac{1}{\sqrt{Q N_{RF}}}\{-1,+1\}$ with equal probability.

From Fig.\,\ref{fig:channel} and Section \ref{sub:channel}, the beamspace channel vectors at different subcarriers, even if different due to beam squint, are correlated through antenna array response vector $\ba(\phi_{l,m})$, which is highly similar to a $2$D natural image.
By stacking $M$ beamspace channel vectors into a matrix, we have the following signal recovery model,
\begin{equation}\label{eq:h_matrix}
  [\bar{\by}_{1},\bar{\by}_{2},\ldots,\bar{\by}_{M} ]=\bar{\bW}[\tilde{\bh}_{1},\tilde{\bh}_{2},\ldots,\tilde{\bh}_{M}] + [\bn_{1}^{\mathrm{eq}},
  \bn_{2}^{\mathrm{eq}},\ldots,\bn_{M}^{\mathrm{eq}}].
\end{equation}
If we regard  beam-frequency matrix $[\tilde{\bh}_{1},\tilde{\bh}_{2},\ldots,\tilde{\bh}_{M}]$ as a $2$D natural image, many compressed image recovery method can be borrowed here for beamspace channel estimation, which enables us to develop a model-driven-DL-based channel estimation network.

Before introducing the {\blc DL-based channel estimation} network, we first obtain a linear transformation model by stacking the $\by_{m}$, $\tilde{\bh}_{m}$ and $\bn_{m}^{\mathrm{eq}}$ into
\begin{equation}\label{eq:linearmodel}
  \by = \bA\bh+\bn,
\end{equation}
where $\by = [\bar{\by}_{1}^{T},\bar{\by}_{2}^{T},\ldots,\bar{\by}_{M}^{T}]$, $\bh = [\tilde{\bh}_{1}^{T},\tilde{\bh}_{2}^{T},\ldots,\tilde{\bh}_{M}^{T}]^{T}$,
$\bn = [(\bn_{1}^{\mathrm{eq}})^{T},(\bn_{2}^{\mathrm{eq}})^{T},\ldots,(\bn_{M}^{\mathrm{eq}})^{T}]^{T}$, and $\bA = (\bI \otimes \bar{\bW})$. We denote $\otimes$ as the matrix Kronecker product. The linear transformation model ($\ref{eq:linearmodel}$) will be utilized in the subsequent discussion.

\section{Unsupervised learning for beamspace channel estimation}\label{without_data}
In this section, we propose a  model-driven unsupervised DL network for wideband beamspace channel estimation, named LDGEC-based channel estimator. As in \cite{Modeldriven18DL}, the network is specially designed by unfolding an iterative algorithm, GEC algorithm, with the DL-based denoiser. After introducing the network architecture and  DnCNN denoiser, we elaborate the SURE loss and the layer-by-layer training approach, which are  critical to implementing LDGEC network with unsupervised learning.
\begin{figure*}[t]
  \centering
  \includegraphics[width=16cm]{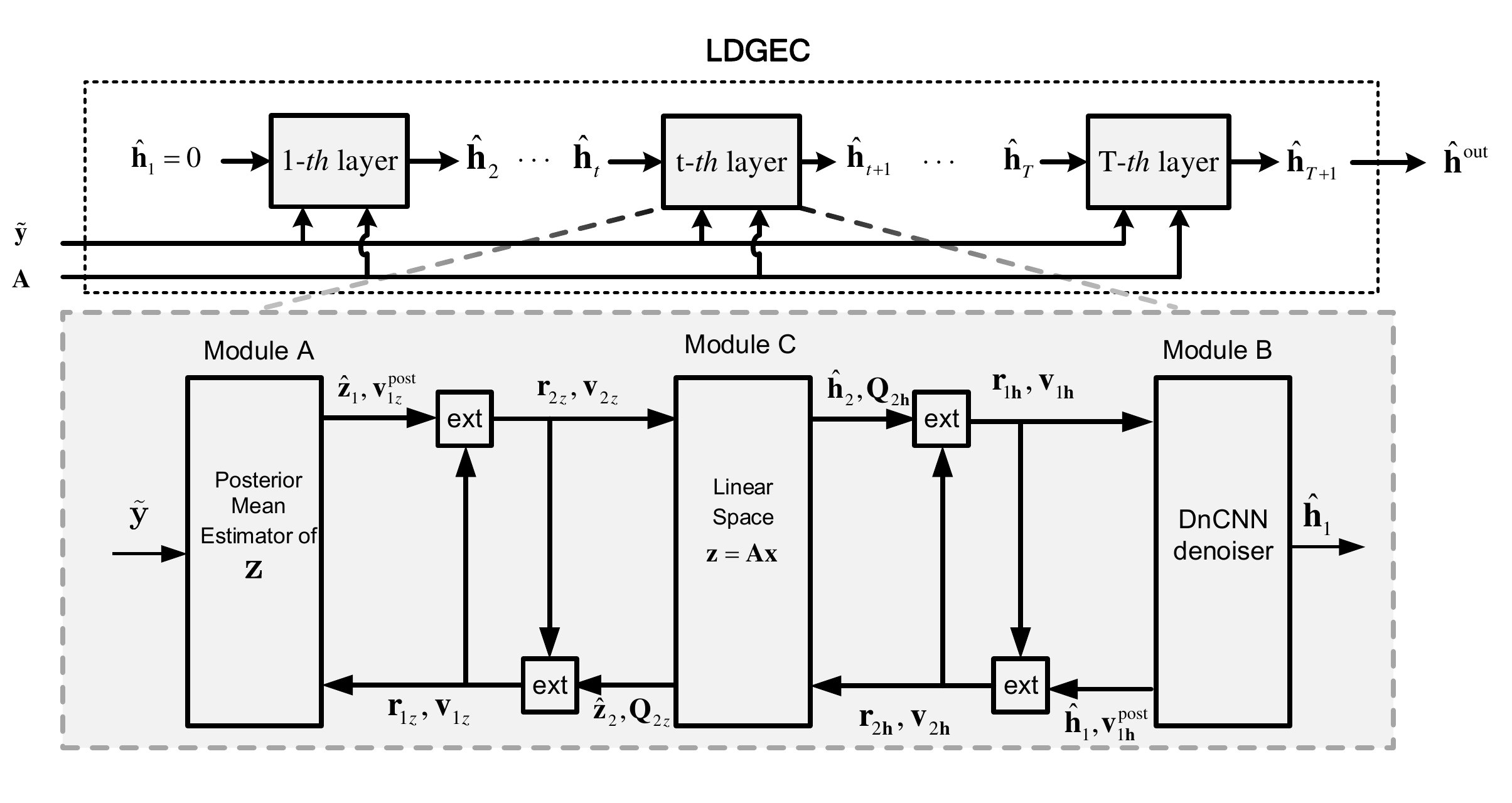}
  \caption{.~~ The network structure of LDGEC-based channel estimator.}\label{fig:LDGEC}
\end{figure*}

\subsection{LDGEC-based channel estimator}
 As illustrated in Fig.\,\ref{fig:LDGEC}, the input of the LDGEC network is the received signal vector, $\tilde{\by}$, and the linear transform  matrix, $\bA$,  while the final output is $\hat{\bh}^{\mathrm{out}}$, the estimated channel vector. The LDGEC  network consists of $T$ layers connected in cascade. We replace the posterior mean estimator in the GEC algorithm with the DnCNN denoiser and deep unfold the GEC algorithm into neural network. Each iteration of the GEC algorithm can be interpreted as each layer of the LDGEC network. As each layer of LDGEC has the same structure except the learnable parameters in DnCNN denoiser, we omit the layer index $t$ in Fig. \ref{fig:LDGEC} and Algorithm 1.

As illustrated in the figure, each layer of the LDGEC network has three modules. Specifically, Module A computes the posterior mean and variance of $\bz = \bA \bh $, module B performs denoising from the noisy signal, $\br_{1\bh}$, by using the advanced DnCNN denoiser, and module C provides the framework that constrains the estimation problem into the linear space $\bz = \bA\bh$.
\begin{algorithm*}\label{algGE}\small{
\caption{LDGEC-based channel estimator} 
\hspace*{0.02in} {\bf Input:} 
 Received signals $\tilde{\mathbf{y}}$, linear transform matrix $\mathbf{A}$, likelihood $\sfP(\tilde{\mathbf{y}}|\mathbf{z})$\\
\hspace*{0.02in} {\bf Output:} 
Recovered signal $\hat{\bh}^{\mathrm{out}}$.\\
\hspace*{0.02in} {\bf Initialize:}
$t \leftarrow 1$, $\mathbf{r}_{1\mathbf{z}}\leftarrow \mathbf{0}$, $\mathbf{r}_{2\mathbf{h}}\leftarrow \mathbf{0}$, $\mathbf{v}_{1\mathbf{z}}\leftarrow P_{z} \mathbf{1}$, and $\mathbf{v}_{2\mathbf{h}}\leftarrow P_{h} \mathbf{1}$.

 \For{$t = 1,\cdots, T$ }{

 \textbf{Module A:}

(1) Compute the posterior mean and covariance of $\mathbf{z}$\\
\nl     $\hat{ \mathbf{z}}_{1} =\mathtt{E}\left\{\mathbf{z}|\mathbf{\mathbf{r}}_{1\mathbf{z}},\mathbf{v}_{1\mathbf{z}} \right\},$\\
\nl    $\mathbf{v}_{1\mathbf{z}}^{\mathrm{post}} = \mathtt{Var} \left\{ \mathbf{z}|\mathbf{\mathbf{r}}_{1\mathbf{z}},\mathbf{v}_{1\mathbf{z}} \right\}.$

(2) Compute the extrinsic information of $\mathbf{z}$\\
 \nl$\mathbf{v}_{2\mathbf{z}}= \mathbf{1}\oslash {\left( \mathbf{1}\oslash\mathbf{v}_{1\mathbf{z}}^{ \mathrm{post}}- \mathbf{1}\oslash\mathbf{v}_{1\mathbf{z}} \right)}, $\\
 \nl$\mathbf{r}_{2\mathbf{z}}= \mathbf{v}_{2\mathbf{z}} \odot {\left( \hat{\mathbf{z}}_{1}\oslash\mathbf{v}_{1\mathbf{z}}^{ \mathrm{post}}-\mathbf{r}_{1\mathbf{z}}\oslash\mathbf{v}_{1\mathbf{z}} \right)}.$

   \textbf{Module C:}

(3) Compute the mean and covariance of $\mathbf{h}$  from the linear space\\
   \nl  $\mathbf{Q}_{2\mathbf{h}}={\left(\mathrm{Diag}(\mathbf{1}\oslash \mathbf{v}_{2\mathbf{h}})+\mathbf{A}^{H}\mathrm{Diag}(\mathbf{1}\oslash\mathbf{v}_{2\mathbf{z}})\mathbf{A} \right)}^{-1}, $ \\
  \nl $\hat{\mathbf{h}}_{2}=\mathbf{Q}_{2\mathbf{h}} \left(\mathbf{r}_{2\mathbf{h}}\oslash\mathbf{v}_{2\mathbf{h}}+\mathbf{A}^{H}\mathbf{r}_{2\mathbf{z}}\oslash\mathbf{v}_{2\mathbf{z}}\right).$

(4) Compute the extrinsic information of $\mathbf{h}$\\
  \nl   $\mathbf{v}_{1\mathbf{h}}=\mathbf{1}\oslash \left( \mathbf{1}\oslash\mathbf{d}(\mathbf{Q}_{2\mathbf{h}}) - \mathbf{1}\oslash\mathbf{v}_{2\mathbf{h}} \right),$\\
  \nl   $\mathbf{r}_{1\mathbf{h}}=\mathbf{v}_{1\mathbf{h}} \odot \left( \hat{\mathbf{h}}_{2}\oslash\mathbf{d}(\mathbf{Q}_{2\mathbf{h}}) - \mathbf{r}_{2\mathbf{h}}\oslash\mathbf{v}_{2\mathbf{h}} \right).$

 \textbf{Module B:}

(5) Compute the mean and covariance of $\mathbf{h}$ \\
\nl
  $\hat{\mathbf{h}}_{1}= D_{\hat{\sigma}}(\mathbf{\mathbf{r}}_{1\mathbf{h}},\mathbf{v}_{1\mathbf{h}})$ \\
\nl
 $\mathbf{v}_{1\mathbf{h}}^{\mathrm{post}}=\frac{1}{MN}\mathrm{div}D_{\hat{\sigma}} (\mathbf{\mathbf{r}}_{1\mathbf{h}})
\mathrm{ avg}(\mathbf{v}_{1\mathbf{h}})$

(6) Compute the extrinsic information of $\mathbf{h}$\\
 \nl    $\mathbf{v}_{2\mathbf{h}}= \mathbf{1}\oslash {\left( \mathbf{1}\oslash\mathbf{v}_{1\mathbf{h}}^{\mathrm{post}}- \mathbf{1}\oslash\mathbf{v}_{1\mathbf{h}} \right)}, $\\
  \nl   $\mathbf{r}_{2\mathbf{h}}=\mathbf{v}_{2\mathbf{h}} \odot {\left( \hat{\mathbf{h}}_{1}\oslash\mathbf{v}_{1\mathbf{h}}^{\mathrm{post}} - \mathbf{r}_{1\mathbf{h}}\oslash\mathbf{v}_{1\mathbf{h}} \right)} . $

 \textbf{Module C:}

(7) Compute the mean and covariance of $\mathbf{z}$ from the linear space\\
 \nl   $\mathbf{Q}_{2\mathbf{h}}={\left(\mathrm{Diag}(\mathbf{1}\oslash \mathbf{v}_{2\mathbf{h}})+\mathbf{A}^{H}\mathrm{Diag}(\mathbf{1}\oslash\mathbf{v}_{2\mathbf{z}})\mathbf{A} \right)}^{-1},$\\
 \nl   $ \hat{\mathbf{h}}_{2}=\mathbf{Q}_{2\mathbf{h}} \left(\mathbf{r}_{2\mathbf{h}}\oslash\mathbf{v}_{2\mathbf{h}}+\mathbf{A}^{H}\mathbf{r}_{2\mathbf{z}}\oslash\mathbf{v}_{2\mathbf{z}}\right),$\\
 \nl   $\mathbf{Q}_{2\mathbf{z}}=\mathbf{A}\mathbf{Q}_{2\mathbf{h}}\mathbf{A}^{H}, $\\
 \nl   $\hat{\mathbf{z}}_{2}=\mathbf{A}\hat{\mathbf{h}}_{2}. $

(8) Compute the extrinsic information of $\mathbf{z}$ \\
\nl  $\mathbf{v}_{1\mathbf{z}}= \mathbf{1}\oslash {\left(\mathbf{1}\oslash\mathbf{d}(\mathbf{Q}_{2\mathbf{z}}) - \mathbf{1}\oslash\mathbf{v}_{2\mathbf{z}} \right)},$ \\
\nl     $\mathbf{r}_{1\mathbf{z}}=\mathbf{v}_{1\mathbf{z}} \odot {\left( \hat{\mathbf{z}}_{2}\oslash\mathbf{d}(\mathbf{Q}_{2\mathbf{z}}) - \mathbf{r}_{2\mathbf{z}} \oslash \mathbf{v}_{2\mathbf{z}} \right)}. $
     }
     }
\end{algorithm*}
Modules A, B, and C are executed iteratively, as in the figure. In addition, each module uses the turbo principle as in iterative decoding; that is, each module passes the extrinsic messages to its next module. The three modules are executed iteratively until convergence or terminated by fixed number of layers.

Before introducing the principle of the LDGEC network, we define two auxiliary variables,
\begin{equation}\label{eq:defPxPz}
  P_{h} = 1~~ \mathrm{and}  P_{z} = P_{h} \cdot \mathsf{tr}(\mathbf{A}^H\mathbf{A})/MN_{RF}Q,
\end{equation}
which are interpreted as the powers of $h_{n}$ and $z_{n}$, respectively. $h_{n}$  and $z_{n}$ denote the $n$-th
element in $\bh$ and $\bz$, respectively. The $P_{h}$ and $P_{z}$ are important for network initialization. The algorithm for the LDGEC-based channel estimator is listed in Algorithm 1.

To better understand the LDGEC network, we provide detailed explanations. Lines 1--2 compute the posterior mean and variance of $z_{n}$ from quantized measurements $\tilde{y}_{n}$, and the expectation w.r.t. the posterior
\begin{equation}\label{posterior_estimate_z}
  \sfP_{Z}(z_{n}|\tilde{y}_{n})=\frac{\sfP_{\mathrm{out}}(\tilde{y}_{n}|z_{n})\sfP_{Z}(z_{n})}{\int\sfP_{\mathrm{out}}(\tilde{y}_{n}|z_{n})\sfP_{Z}(z_{n})dz_{n}},
\end{equation}
where $\sfP_{Z}(z_{n})$ is assumed to be $\cN_{\bbC}(z_{n};r_{1z,n},v_{1z,n})$.
To clearly understand Lines 1 and 2 in Algorithm 1, we take the quantized and unquantized channels as two examples.

\textbf{Unquantized channel}: If with infinite-resolution ADCs, the received signal at the BS, $\tilde{\by}=\by$ and the posterior probability $\sfP_{\mathrm{out}}(\tilde{y}_{n}|z_{n})$ is given by

\begin{equation}\label{eq:posterior_probabi}
  \sfP_{\mathrm{out}}(\tilde{y}_{n}|z_{n}) = \frac{1}{\pi \sigma_{n}^{2}}e^{|\tilde{y}_{n}-z_{n}|/\sigma_{n}^2}.
\end{equation}

Thus, the explicit expressions of the posterior mean and variance will be
\begin{align}
    \hat{z}_{1}
   &= r_{1z} + \frac{v_{1z}}{v_{1z}+\sigma_{n}^{2}}(\tilde{y}-r_{1z}),
   \label{finiteADCmean} \\
    v_{1z}^{\mathrm{post}} & =v_{1z}-\frac{v_{1z}^{2}}{v_{1z}+\sigma_{n}^{2}} ,
     \label{finiteADCvar}
\end{align}

\textbf{Quantized channel}: If the low-resolution ADCs are used in the BS, the received signal $\tilde{\by}=\cQ_{c}(\by)$, where $\cQ_{c}$ is the complex-valued quantizer. Then, the explicit expressions of the posterior mean and variance can be derived similar to \cite[Appendix A]{CKWen2016TSP} as
\begin{align}
    \hat{z}_{1}
   &= r_{1z} + \frac{\sign(\tilde{y}) v_{1z} }{\sqrt{2(\sigma_{n}^2 + v_{1z})}} \left( \frac{\phi(\eta_1)-\phi(\eta_2)}{\Phi(\eta_1)-\Phi(\eta_2)} \right),
   \label{eq:hatZ_RealGaussian} \\
    v_{1z}^{\mathrm{post}} & = \frac{v_{1z}}{2} - \frac{(v_{1z})^2}{2(\sigma_{n}^2 + v_{1z})}\times  \nonumber \\
     & \left( \frac{\eta_1\phi(\eta_1)-\eta_2\phi(\eta_2)}{\Phi(\eta_1)-\Phi(\eta_2)}
     + \left(\frac{\phi(\eta_1)-\phi(\eta_2)}{\Phi(\eta_1)-\Phi(\eta_2)}\right)^2 \right),
     \label{eq:mseZ_RealGaussian}
\end{align}
where
\begin{subequations} \label{eq:eta_def}
\begin{align}
    \eta_1 &= \frac{\sign(\tilde{y})r_{1z}-\min\{|r^{\mathrm{low}}|,|r^{\mathrm{up}}|\}}{\sqrt{(\sigma_{n}^2 + v_{1z})/2}}, \\
    \eta_2 &= \frac{\sign(\tilde{y})r_{1z}-\max\{|r^{\mathrm{low}}|,|r^{\mathrm{up}}|\}}{\sqrt{(\sigma_{n}^2 + v_{1z})/2}},
\end{align}
\end{subequations}
where $r^{\mathrm{low}}$ and $r^{\mathrm{up}}$ are  the lower and upper thresholds associated with $\tilde{y}_{n}$, respectively. For notational convenience, we omit index $n$ and have
\begin{subequations}\label{quanbound}
    \begin{equation}
    r^{\mathrm{low}} = \left\{ \begin{array}{ll}
    \tilde{y}-\frac{\Delta}{2}, & \textrm{for $\tilde{y}\ge -{\left(\frac{2^{\kappa}}{2}-1\right)}\Delta$},\\
    -\infty, & \textrm{otherwise},
    \end{array} \right.
    \end{equation}
    and
    \begin{equation}
    r^{\mathrm{up}} = \left\{ \begin{array}{ll}
    \tilde{y}+\frac{\Delta}{2}, & \textrm{for $\tilde{y}\le {\left(\frac{2^{\kappa}}{2}-1\right)}\Delta$},\\
    \infty, & \textrm{otherwise}.
    \end{array} \right.
    \end{equation}
\end{subequations}
In this paper, we mainly focus on  a typical uniform midrise quantizer with quantization step size $\Delta$. It maps a real-valued input into the nearest value in
\begin{equation} \label{2.2}
    \cR_{{\kappa}} \triangleq {\left\{ \Big({-\frac{1}{2}}+b\Big) \Delta; \,\, b=-\frac{2^{{\kappa}}}{2}+1, \cdots, \frac{2^{{\kappa}}}{2} \right\}} ,
\end{equation}
where $\kappa$ is the quantization bits.

The real and imaginary parts are quantized separately, and each complex-valued channel can be decoupled into two real-valued channels. Expressions (\ref{eq:hatZ_RealGaussian}) and (\ref{eq:mseZ_RealGaussian}) are the estimators only for the real part of $\hat{z}_{1}$. To facilitate notation, we have
abused $\tilde{y}$ and $\hat{z}_{1}$ in (\ref{eq:hatZ_RealGaussian}) and (\ref{eq:mseZ_RealGaussian}) to denote $ \mathrm{Re}(\tilde{y})$ and $ \mathrm{Re}(\hat{z}_{1})$, respectively, and we omit index $n$ in the aforementioned expression. The estimator for the imaginary part $\mathrm{Im}(\hat{z}_{1})$ can be obtained similarly as (\ref{eq:hatZ_RealGaussian}) and (\ref{eq:mseZ_RealGaussian}) while $\tilde{y}$ and $b$ should be replaced by
$\mathrm{Im}(\tilde{y})$ and $b'$, respectively.

Lines $3$--$4$ compute the extrinsic information of $\bz$ using the turbo principle. Lines $5$--$6$ perform the linear  minimum mean-squared error (LMMSE) estimate of $\bh$ under the following assumption,
\begin{equation}\label{LMMSE_estimate}
  \br_{2\bz}=\bz_{2}+\bw_{2\bz},
\end{equation}
where $\bw_{2\bz} \sim \cN_{\bbC}(\mathbf{0},\mathrm{Diag}(\bv_{2\bz}))$, $\bz_{2}=\bA\bh_{2}$, and $\bh_{2} \sim \cN_{\bbC}(\bh_{2};\br_{2\bh},\mathrm{Diag}(\bv_{2\bh}))$. Lines $7$--$8$ compute the extrinsic information of $\bh$ and pass it to module B as a prior information. Lines $9$--$10$ estimate the mean, $\hat{\bh}_{1}$, and variance, $\bv_{1\bh}^{\mathrm{post}}$ by considering the true prior $\sfP(\mathbf{h})$, which is assumed to estimate $\bh$ from several AWGN observations, that is,
\begin{equation}\label{AWGN_estimate}
\br_{1\bh}=\bh+\bw_{1\bh},
\end{equation}
where $\bw_{1\bh} \sim \cN_{\bbC}(\mathbf{0},\mathrm{Diag}(\bv_{1\bh}))$. As  channel $\bh$ can be regarded as a $2$D natural image, we utilize the advanced DnCNN denoiser \cite{DnCNN2017Zhang} in Lines  $9$--$10$ to recover channel $\bh$ from equivalent noisy observations $\br_{1\bh}$. Lines $11$--$12$ compute the extrinsic information of $\bh$ using the turbo principle, and lines $13$--$16$ constrain the estimated problem into a linear space $\bz=\bA\bh$ which performs the same procedure as Lines $5$--$6$. Lines $17$--$18$ compute the extrinsic information of $\bz$ and pass to module A as the prior information.

Posterior variance $\mathbf{v}_{1\mathbf{h}}^{\mathrm{post}}$ is determined by $\mathrm{div} D_{\hat{\sigma}^t}(\br_{1\bh})\mathrm{avg}(\bv_{1\bh})$, where the divergence $\mathrm{div} D_{\hat{\sigma}^t}(\br_{1\bh})$  is simply the sum of the partial derivates with respect to each element of $\br_{1\bh}$. It can be expressed by
\begin{equation}\label{eq:divD}
  \mathrm{div} D_{\hat{\sigma}^t}(\br_{1\bh}) = \sum_{i=1}^{n}\frac{\partial D_{\hat{\sigma}^t}(\br_{1\bh})}{\partial r_{1h,i}},
\end{equation}
where $r_{1h,i}$ is the $i$-th element of $\br_{1\bh}$. Although simple denoisers often yield a closed form for their divergence, high-performance denoisers are often data-dependent; making it very difficult to characterize their input-output relationship explicitly for most DL-based denoisers. Therefore, we  calculate a good approximation for the divergence.

We use the following Monte-Carlo approximation to compute divergence $\mathrm{ div} D_{\hat{\sigma}^t}(\cdot)$.
Using an independent and identically distributed (i.i.d.) random vector $\bb \sim \mathcal{N}(\mathbf{0},\bI)$, we can estimate the divergence with
{\setlength{\arraycolsep}{0.0em}
\begin{eqnarray}
\label{eqn:div_est}
\mathrm {div} D_{\hat{\sigma}^t}&=&\lim\limits_{\epsilon\rightarrow 0}\mathbb{E}_{\bb} \left\{\bb^{T}\left(\frac{D_{\hat{\sigma}^t}(\br_{1\bh}+\epsilon \bb)-D_{\hat{\sigma}^t}(\br_{1\bh})}{\epsilon}\right)\right\}\\ &\approx &  \frac{1}{\epsilon}\bb^{T}(D_{\hat{\sigma}^t}(\br_{1\bh}+\epsilon \bb)-D_{\hat{\sigma}^t}(\br_{1\bh})),
\end{eqnarray}\setlength{\arraycolsep}{5pt}}

where $\epsilon=\|\br_{1\bh}\|_\infty/1000$ is an arbitrary small number. Equation ($\ref{eqn:div_est}$) is originated from the law of large numbers. The expectation can be approximated with Monte Carlo sampling and a single sample can well approximate the expectation.

To improve robustness, we use an auto-regressive filter to smooth the update of ($\bv_{1\bz},\br_{1\bz}$) by
\begin{equation}\label{eq:dampingv}
  \bv_{1\bz}^{t+1} = \beta\cdot\mathbf{1}\oslash {\left(\mathbf{1}\oslash\mathbf{d}(\mathbf{Q}_{2\mathbf{z}}^{t+1}) - \mathbf{1}\oslash\mathbf{v}_{2\mathbf{z}}^{t+1} \right)}+(1-\beta)\bv_{1\bz}^{t},
\end{equation}
\begin{equation}\label{eq:dampingr}
\br_{1\bz}^{t+1} = \beta \cdot\bv_{1\bz}^{t+1}\oslash {\left(\mathbf{1}\oslash\mathbf{d}(\mathbf{Q}_{2\mathbf{z}}^{t+1}) - \mathbf{1}\oslash\mathbf{v}_{2\mathbf{z}}^{t+1} \right)}+(1-\beta)\br_{1\bz}^{t},
\end{equation}
where a small $\beta$ is the damping factor. Furthermore, a small constant threshold $\epsilon_{1}=5\times e^{-7}$ should be set to restrict the minimum variance allowed per iteration and avoid numerical instabilities, that is, $\bv_{1\bz}$ = max($\epsilon_{1}$, $\bv_{1\bz}$) and $\mathbf{v}_{1\mathbf{h}}^{\mathrm{post}}$=max($\epsilon_{1}$, $\mathbf{v}_{1\mathbf{h}}^{\mathrm{post}}$).
\subsection{DnCNN denoiser}
\begin{figure*}[t]
  \centering
  \includegraphics[width=16cm]{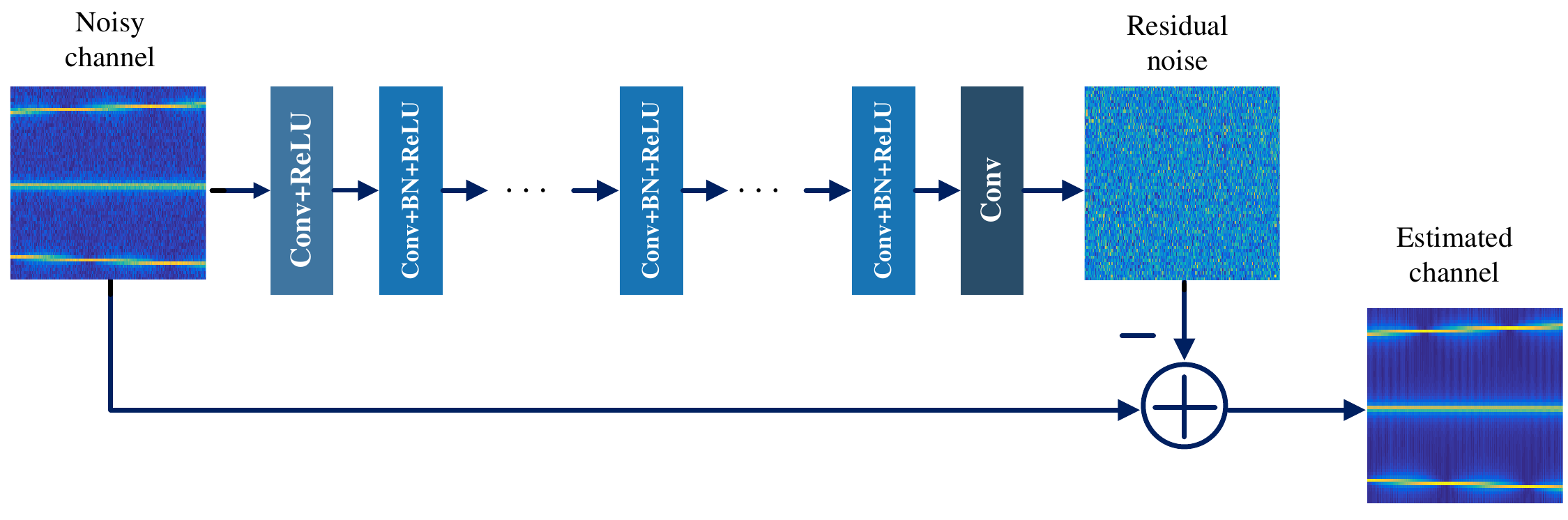}
  \caption{.~~Network architecture of the DnCNN denoiser.}\label{fig:DnCNN}
\end{figure*}
The denoiser used in the LDGEC network plays a key role in channel estimation. {\blc We consider the state-of-the-art DnCNN denoiser\footnote{Although several new denoisers have been proposed for image denoising problem recently \cite{denoiser2018overview}, they have similar performance with DnCNN denoiser.}}. The DnCNN is first proposed in \cite{DnCNN2017Zhang} to handle the Gaussian denoising problem with an unknown noise level, which is more accurate and  faster than competing techniques. {\blc Fig. \ref{fig:DnCNN} illustrates the network architecture of the DnCNN denoiser. It consists of $20$ convolutional layers. The first convolutional layer uses $64$ different $3\times 3\times 1$ filters and is followed by a rectified linear unit (ReLU). Each of the succeeding $18$ convolutional layers uses $64$ different $3\times3\times64$ filters, each followed by batch-normalization and a ReLU. The final convolutional layer uses one separate $3\times 3\times 64$ filters to reconstruct the signal. Instead of learning a mapping directly from a noisy image to a denoised image, learning the residual noise is beneficial.}

We plot three pseudo-color images of noisy channel, residual noise, and estimated channel in Fig. \ref{fig:DnCNN}. The network is given the noisy observation $\bh + \hat{\sigma} \bw$ as an input, where $\bw$ is the AWGN and noise variance $\hat{\sigma}$ is uniformly generated from a specific interval. The network produces residual noise $\hat{\bz}$, rather than an estimated channel $\hat{\bh}$, as an output. This method, known as residual learning \cite{RL2016He}, renders the network to remove the highly structured natural image rather than the unstructured noise. Consequently, residual learning improves both the training times and accuracy of a network. Furthermore, the DnCNN adopts the method of batch normalization, can speed up the training process, and boost the denoising performance \cite{DnCNN2017Zhang}.

\subsection{Stein’s unbiased risk estimator}
Recently, many DL-based channel estimators have been proposed for different communication scenarios \cite{DL2018HE}. A common limitation of these works is the extensive real channel data should be obtained before training the network because the MSE function $\mathrm{MSE} = \mathbb{E}\|\hat{\bh}-\bh\|$ involves in the real channel data $\bh$. These requirements bring significant challenges when the real data cannot be obtained, e.g., the DL-based channel estimator is equipped in a new channel environment and only received signal $\tilde{\by}$ is obtained at the BS. {\blc Furthermore, the existing DL-based channel estimator utilizes supervised way and requires a large number of real channel data, which defeats the point of channel estimation. In this circumstance, how to train the network without the real channel data and only with  measurements corresponding to the pilot symbols is significantly important. To solve the problem, we introduce the SURE loss function \cite{SURE1981}, which is a classical approach for learning images from noisy observations  and has been extended to linear noisy measurements. It has been applied in medical imaging, microscopy, and astronomy, where the ground truth data is rarely available \cite{SURE2018Metzler}. The SURE loss is an unbiased estimator of
\begin{equation}\label{equnbiased}
    \mathrm{MSE} = \mathbb{E}_{\bw} [\frac{1}{P}\|\bh - f(\by_{\bw})\|^{2}],
\end{equation}
where $\by_{\bw} = \bh+\bw$ and $f(\cdot)$ is an estimator of $\bh$ from $\by_{\bw}$.
Therefore, it can be used for training a DL-based denoiser to replace the MSE loss. }

The goal of  channel estimation is to reconstruct a channel $\bh$ from  noisy linear observations $\by = \bA\bh + \bn$ with the {\blc known} linear transform matrix $\bA$. We are given training measurements $\by^{1}$, $\by^{2}$, \ldots, $\by^{D}$ but not the real channel $\bh^{1}$, $\bh^{2}$, \ldots, $\bh^{D}$. Without access to $\bh^{1}$, $\bh^{2}$,\ldots, $\bh^{D}$, the ground truth data, we cannot train the DnCNN denoiser by minimizing the traditional MSE loss function. Fortunately, we can use the SURE loss instead. SURE is a model selection technique first proposed by its namesake in \cite{SURE1981}. It provides an unbiased estimate of the MSE {\blc for} an estimator of the mean of a Gaussian distributed random vector with an unknown mean. Let $\bx$ denote a vector we would like to estimate from noisy observations $\br_{1\bh} = \bh + \bw_{1\bh}$ where $\bw_{1\bh} \sim \mathcal{N}_{\mathbb{C}}(\boldsymbol{0},\mathrm{Diag}(\bw_{1\bh}))$. We assume the DnCNN function $f_{\bm{\theta}}(·)$ is a weakly differentiable function parameterized by $\bm{\theta}$, which receives noisy observations $\br_{1\bh}$ as input and provides an estimate of $\bh$ as output. Then, according to \cite{SURE1981,SURE2018Metzler}, we can express the expectation of the MSE of the real channel $\bh$ and equivalent noisy observations  $\br_{1\bh}$ with respect to the
random variable $\bw_{1\bh}$ as follows,
\begin{equation}\label{eq:MSE}
  \mathrm{MSE} = \mathbb{E}_{\bw_{1\bh}} [\frac{1}{P}\|\bh - f_{\bm{\bm{\theta}}}(\br_{1\bh})\|^{2}].
\end{equation}
Then, the MSE loss can be computed as follows,
\begin{align}\label{eq:sure}
  \mathbb{E}_{\bw_{1\bh}} [\frac{1}{P}\|\bh - f_{\bm{\theta}}(\br_{1\bh})\|^{2}] &= \mathbb{E}_{\bw_{1\bh}} [\frac{1}{P}\|\br_{1\bh} - f_{\bm{\theta}}(\br_{1\bh})\|^{2}] -v_{1h}^{2} \\  \nonumber
  &+ \frac{2v_{1h}^{2}}{P}\mathrm{div}(f_{\bm{\theta}}(\br_{1\bh})),
\end{align}
where $P = MN$ and $\mathrm{div}(\cdot)$ stands for divergence defined as ($\ref{eq:divD}$).
Note that two terms within the SURE loss depend on  parameter $\bm{\theta}$. The first term, $\mathbb{E}_{\bw_{1\bh}} [\frac{1}{P}\|\br_{1\bh} - f_{\bm{\theta}}(\br_{1\bh})\|^{2}]$, indicates the difference between estimate $f_{\bm{\theta}}(\br_{1\bh})$ and
observation $\br_{1\bh}$ (bias). The second term, $\frac{2v_{1h}^{2}}{P}\mathrm{div}(f_{\bm{\theta}}(\br_{1\bh}))$, penalizes the denoiser for varying as the input is changed. Thus, SURE is a natural way to control the trade-off between the bias and variance of a recovery algorithm.

The critical challenge {\blc for} using SURE in practice is to compute  divergence $\mathrm{div}(f_{\bm{\theta}}(\br_{1\bh}))$. For the advanced DnCNN denoiser, the divergence is hard or even impossible to express analytically. Therefore, we cannot obtain a closed form for the divergence. Similar to (\ref{eqn:div_est}), we can use a Montel Carlo method to estimate the divergence $\mathrm{div}(f_{\bm{\theta}}(\br_{1\bh}))$. Combining the SURE loss in (\ref{eq:sure}) and the estimate of divergence $\mathrm{div}(f_{\bm{\theta}}(\br_{1\bh}))$, we can minimize the  MSE loss function of a denoising problem without ground truth data. Note that minimizing SURE loss requires propagating gradients with respect to the Monte Carlo estimate of  divergence (\ref{eq:divD}). Although the gradients are challenging to compute by hand, we can resort to TensorFlow's auto-differentiation capabilities to propagate it.

\subsection{Layer-by-layer training}
A significant reason for the LDGEC-based channel estimator  trained with SURE loss is layer-by-layer training. From (\ref{eq:sure}), the computational process of the SURE loss requires noisy observations $\br_{1\bh}$ and equivalent noise variance $v_{1h}$. 
This method takes advantage of the fact that each layer of the LDGEC-based channel estimator is to solve a denoising problem with known variance $v_{1h}$ and noisy observations $\br_{1\bh}$. As the LDGEC network can decouple the linear model in (\ref{eq:linearmodel}) into several equivalent AWGN models (\ref{AWGN_estimate}) in each layer and $\bv_{1\bh}$ computed in line 7 in Algorithm 1 is accurate enough to describe the variance of  $\bw_{1\bh}$. Therefore, we can train the $t$-th layer LDGEC network with the SURE loss, estimated variance $\bv_{1\bh}^{t}$ and noisy observations $\br_{1\bh}^{t}$ through layer-by-layer training.

In the $t$-th round of the layer-by-layer training, the loss function is given by
\begin{align}\label{eq:cost}
   L^{t}_{\mathrm{SURE}}(\bm{\theta}^{t}) & = \mathop{\arg\min}_{\bm{\theta}^{t}} \sum_{d=1}^{D} [\frac{1}{P}\|\br_{1\bh}^{t,d} - f_{\bm{\theta}}(\br_{1\bh}^{t,d})\|^{2}] -v_{1h}^{2} \\ \nonumber
   &+ \frac{2v_{1h}^{2}}{P}\mathrm{div}(f_{\bm{\theta}}(\br_{1\bh}^{t,d})),
\end{align}
where $D$ is the number of mini-batches in the $t$-th round and $\br_{1\bh}^{t,d}$ is the corresponding noisy observations for sample $\bh^{d}$ in the $l$-th LDGEC network. $\bm{\theta}^{t}$ is the required learnable variables in the $t$-th layer. After training the first to $t$-th layers, a new $t + 1$ layer is appended to the LDGEC network and the entire network is trained again for $D$ mini-batches. Although the objective function is changed, the values of the variables $\bm{\theta}^{0}$, \ldots, $\bm{\theta}^{t-1}$ of the previous round are taken as the initial ones in the optimization process for the new round. In summary, the layer-by-layer training updates  variables $\bm{\theta}^{t}$ in a sequential manner from the first layer to the last layer.

{\blc
\subsection{Complexity analysis}

The computational complexity required for the LDGEC network in each layer is dominant by matrix inverse in lines $5$ and $13$ in Algorithm 1. Generally, the computational complexity of matrix inverse is $\mathcal{O}((MN)^{3})$, which cannot be acceptable. As the matrix $\bA$ is a block diagonal matrix and can be expressed as
\begin{equation}\label{eqblock}
  \mathbf{A}= (\bI \otimes \bar{\bW}) = \left[\begin{array}{c c c}
\mathbf{\bar{\bW}}  \\
 &\ddots   \\
 &&\mathbf{\bar{\bW}}
\end{array}\right].
\end{equation}
 Therefore, the matrix inverse in lines $5$ and $13$ in  algorithm 1 can be computed by matrix inverse with respect to matrix $\bar{\bW}$. By inversion of the partitioned matrix, the total complexity can be reduced  from $\mathcal{O}((MN)^{3})$ to $\mathcal{O}(MN^{3})$. We compare the complexity of LDGEC network with other CS-based algorithms.  As shown in  Table I, the computational complexity of LDGEC is $\mathcal{O}(MN^{3})$, while SSD and BEACHES algorithms have the complexity of $\mathcal{O}(MN_{RF}QL^{2}\Omega^2)$ and $\mathcal{O}(MNlog(N))$, respectively. Furthermore, the computational complexity of OMP is $\mathcal{O}(MN_{RF}QL^{2}\Omega^2)$. Generally, $\Omega $ is the beamspace windows  and assumed  as $\Omega = 4$ when $N=256$ \cite{beam2019Gao}, and is much smaller than $N$. 
Although LDGEC network has higher computational complexity than CS-based algorithms but can achieve better performance in simulation results.
\begin{table}[tp]
	\centering
	\renewcommand{\arraystretch}{1.1}
	\begin{minipage}[c]{1\columnwidth}
		\centering
		\caption{ Complexity of different channel estimators}
		\label{tbl:runtimes}
		\begin{tabular}{@{}lcccc@{}}
			\toprule
			Estimators & LDGEC & SSD & BEACHES & OMP \\
			\midrule
			 Complexity  &$\mathcal{O}(MN^3)$ &$\mathcal{O}(MN_{RF}QL^{2}\Omega^2)$ &$\mathcal{O}(MNlog(N))$  &$\mathcal{O}(NMN_{RF}QL\Omega)$ \\						
			\bottomrule
		\end{tabular}
	\end{minipage}
\end{table}

}

\section{Supervised Learning for Beamspace channel estimation}\label{with_data}
In this section, we investigate the LDGEC-based channel estimator for the mmWave beamspace MIMO systems with supervised learning. After presenting three training methods for the LDGEC network, we briefly introduce the denoiser-by-denoiser training.

\begin{figure*}[t]
  \centering
  \includegraphics[width=16cm]{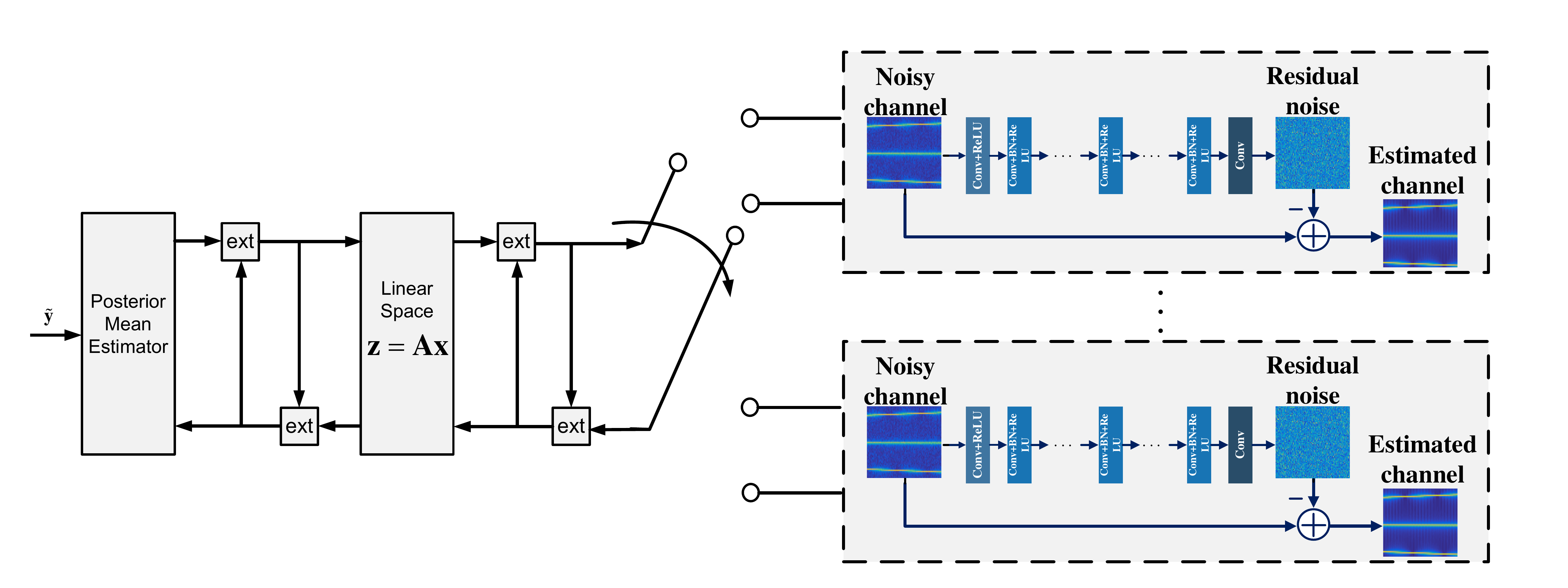}
  \caption{.~~The LDGEC-based channel estimator with denoiser-by-denoiser training.}\label{fig:LDGEC_denoiser}
\end{figure*}

\subsection{Denoiser-by-denoiser training}
 The LDGEC-based channel estimator can also be trained with supervised learning. We introduce three supervised training methods for the LDGEC-based channel estimator with supervised learning as follows,
\begin{itemize}
  \item \emph{End-to-end training}: We can train all the weights of the $T$ layer LDGEC network simultaneously end-to-end. This is the standard method for training a neural network but with high training complexity.
  \item \emph{Layer-by-layer training}: We can train the LDGEC with layer-by-layer {\blc way} by utilizing the MSE loss in (\ref{eq:MSE}). The training process is introduced in Section \ref{without_data}.
  \item \emph{Denoiser-by-denoiser training}: We can decouple the denoisers from the rest of the network
and train the AWGN denoising problems at different noise levels.
\end{itemize}

The principle of the \emph{denoiser-by-denoiser} is to train the DnCNN denoiser for the denoising problem solely instead of including the whole GEC algorithm into network training, {\blc thereby reducing the computational time in the training stage.} %
Note that the DnCNN denoiser in the LDGEC network is trained for a specific noise level interval.
As equivalent noise variance $\hat{\sigma}_{t}^{2}= \mathrm{avg}(\bv_{1\bh})$ is different for each layer in the LDGEC, we need to deploy different DnCNN denoisers for different layers. To address the issue, we decouple the denoisers from the rest of the network and train each on an AWGN denoising problem at different noise levels. In particular, we scale noise level $\hat{\sigma}^{2}$ by multiplying 255 as $\bar{\sigma}^{2} = 255\hat{\sigma}^{2}$ and divide $\bar{\sigma}^{2}$ into intervals [0,10), [10,20), [20,40), [40,60), [60,80), [80,100), [100,150), [150,300), [300,500). For each noise interval, we generate noise variance $\bar{\sigma}^{2}$  uniformly and train a corresponding DnCNN denoiser.

After training the DnCNN denoiser, we deploy the trained DnCNN denoiser into the LDGEC network to perform channel estimation. As illustrated in Fig.\,\ref{fig:LDGEC_denoiser}, we use a selector to choose the corresponding DnCNN denoiser according to the equivalent noise variance $\hat{\sigma}_{t}$ for each layer, e.g.,  we use the denoiser for noise standard deviations between 40 and 60, if $\hat{\sigma}_{t}^{2}*255=55$.

\subsection{MMSE optimal performance}

In \cite{LDAMP2017}, the layer-by-layer and
denoiser-by-denoiser training for LDAMP are proven to achieve MMSE optimal performance under the following three conditions are satisfied,
\begin{itemize}
  \item The elements of  matrix $\bA$ are i.i.d. Gaussian (or sub-gaussian) with zero mean and standard deviation $1/M_{1}$, where $M_{1}$ is the number of rows of $\bA$.
  \item The noise, $\bn$, is also i.i.d. Gaussian.
  \item The denoisers, $D_{\hat{\sigma}^t}(\cdot)$, at each layer are \emph{Lipschitz continuous}\footnote{A denoiser is said to be $L$-Lipschitz continuous if for every $\bx_{1}$ and $\bx_{2}$ we have
  $\|D(\bx_{1})-D(\bx_{2})\| \leq L \|\bx_{1}-\bx_{2}\|_{2}^{2}$ and  the Lipschitz continuity of the convolutional denoiser can be ensured by using weight clipping and gradient norm penalization method \cite{GAN17}.}.
\end{itemize}

Even if the theoretical results have proved that the denoiser-by-denoiser training is optimal, the numerical results in \cite{LDAMP2017} show that LDAMP trained with denoiser-by-denoiser performs slightly worse than the end-to-end and layer-by-layer trained networks due to the discretization of the noise levels ignored in our theory. This gap can be reduced by using a finer discretization of the noise levels or by using deeper denoiser networks to handle a range of noise levels. Although matrix $\bA$ in system model (\ref{eq:linearmodel}) is a block diagonal matrix, rather than a Gaussian matrix, we try to use the denoiser-by-denoiser training method for the LDGEC network and show the numerical results in the following section.
%

\section{Numerical Results}\label{simulation}
In this section, we provide numerical results to show the performance of the proposed model-driven DL network for wideband beamspace channel estimation. First, we elaborate on the implementation details. Then, the performances of the LDGEC-based channel estimator trained with denoiser-by-denoiser and layer-by-layer are presented. Finally, we investigate the performance of the LDGEC-based channel estimator with a reduced number of RF chains and low-resolution ADCs.

\subsection{Implementation details}
{\blc
\begin{table*}[t]
	\centering
\begin{tabular}{|c|c|}
  \hline
  \hline
  Simulation parameters & Value \\
  \hline
  Number of Paths (L) & 3 \\
  \hline
  Number of antennas (N)  & 32 \\
  \hline
  Number of RF chains ($N_{RF}$) & 8 \\
  \hline
  Carrier frequency ($f_{c}$) & 28 GHz \\
  \hline
  Bandwidth ($f_{s}$) & 4 GHz \\
  \hline
  Number of subcarriers (M) & 64, 128 \\
  \hline
  Complex gain ($\beta_{l}$) & $\mathcal{N}_{\mathbb{C}}(0,1)$ \\
  \hline
  Angle ($\theta_{l}$) & $\mathcal{U}(-\pi/2, \pi/2$) \\
  \hline
  Maximum Delay ($\tau_{\mathrm{max}}$) & 20 ns \\
  \hline
  Delay ($\tau_{l}$) & $\mathcal{U}$(0, $\tau_{\mathrm{max}}$)
  \\
  \hline
\end{tabular}
\caption{Simulation parameters}
\label{Table:simulation results}
\end{table*}
{\blc The simulation parameters are listed in Table \ref{Table:simulation results}.} We use the normalized MSE (NMSE) to quantify the accuracy of channel estimation for each user, which is defined as
\begin{equation}\label{eqNMSE}
  \mathrm{NMSE}=\mathbb{E}\left\{\|\hat{\bh}^{\mathrm{out}}-\bh\|_{2}^2/\|\bh\|_{2}^2\right \},
\end{equation}

In our simulation, the DL-based channel estimation network is implemented in Tensorflow by using a PC with GPU NVIDIA GeForce GTX 1080 Ti. The training, validation, and testing sets contain $19200$, $6400$, and $12800$ samples, respectively, and are obtained from the Saleh-Valenzuela channel model in (\ref{eq:h}). The
batch size is set to 16, and epoch is equal to 50. We generate the same adaptive selection network, $\bW$, for the channel sample in each batch,  which is generated independently for different batches. The LDGEC network is trained using the stochastic gradient descent method and the Adam optimizer. The training rate is set to be $0.001$ initially and then dropped to $0.0001$.  As existing deep learning APIs are mostly devoted to processing the real-valued data, we consider equivalent real-valued representation for the system model in (\ref{eq:linearmodel}). We set damping factor $\beta=0.8$ except addition notes. The code will be available at https://github.com/hehengtao/LDGEC.
}
\subsection{Convergence analysis}

\begin{figure}
\begin{minipage}{3in}
  \centerline{\includegraphics[width=3.0in]{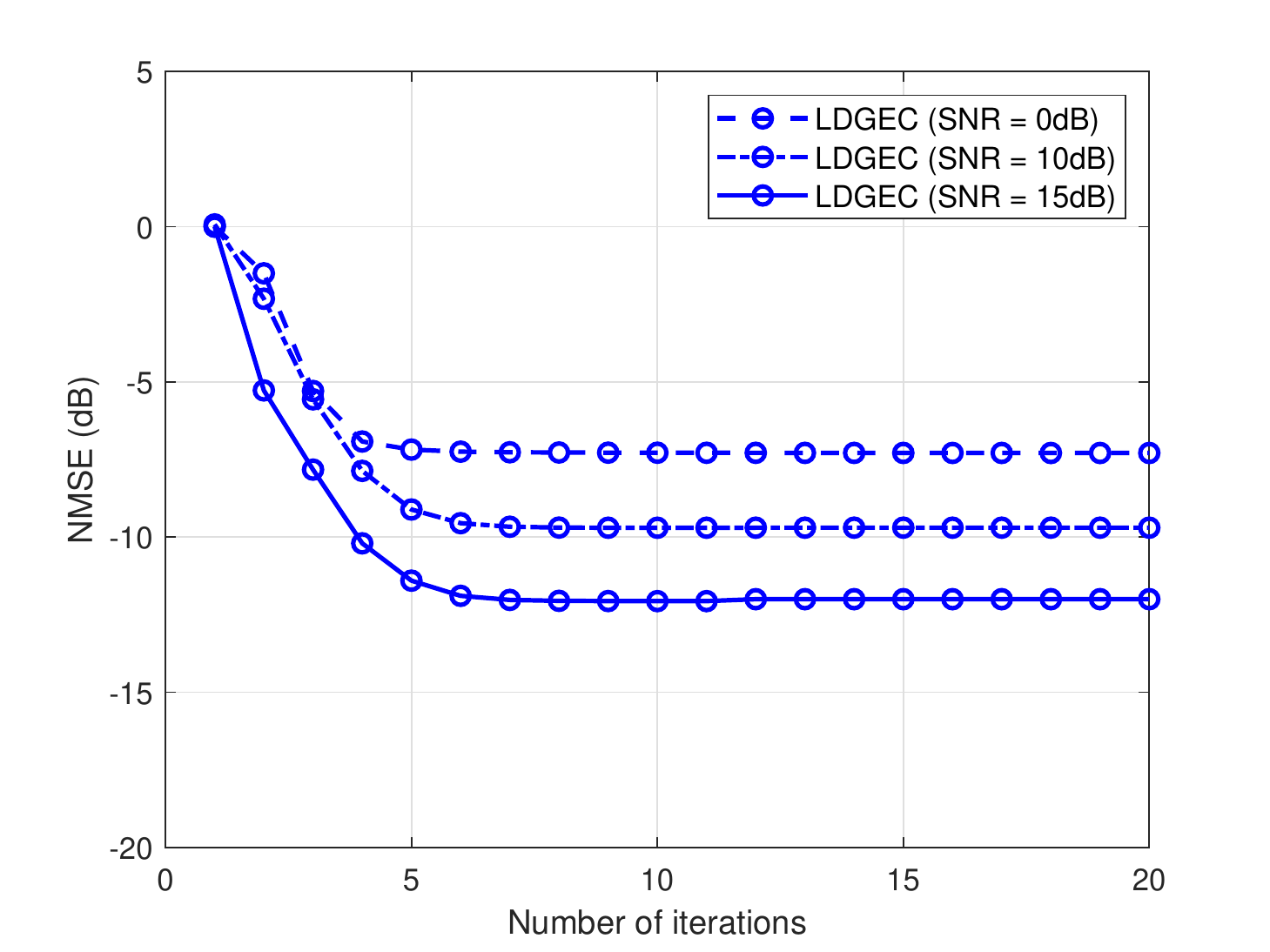}}
  \centerline{(a) Denoiser-by-denoiser training}\label{fig:channela}
\end{minipage}
\hfill
\begin{minipage}{3in}
  \centerline{\includegraphics[width=3.0in]{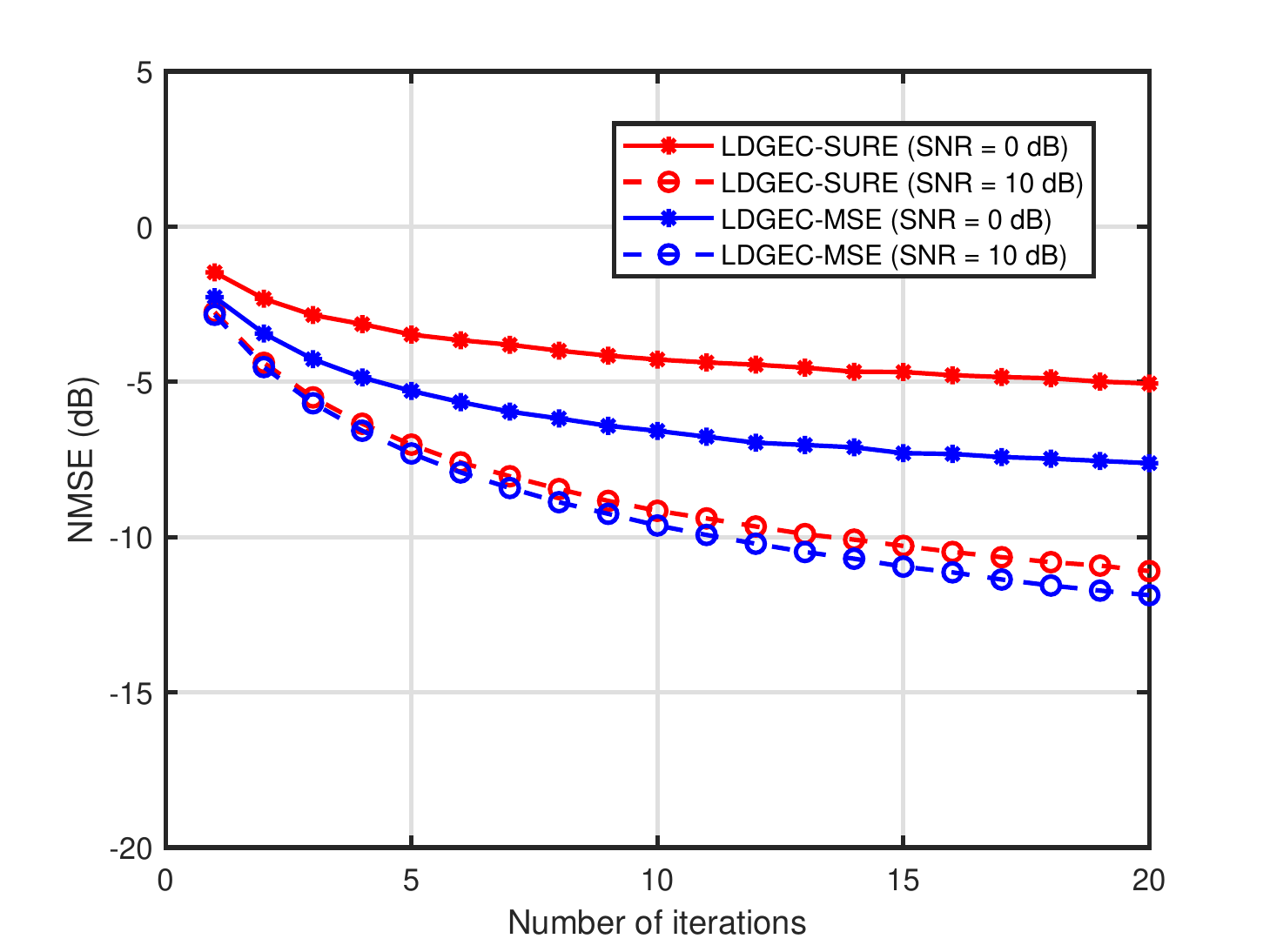}}
  \centerline{(b) Layer-by-layer training} \label{fig:SURE_sim}
\end{minipage}
\caption{.~~Convergence analysis of the LDGEC network.}
\label{fig:LDGEC_iterations}
\end{figure}

\emph{1) With real channel data}: Fig. \ref{fig:LDGEC_iterations}(a) investigates the convergence of the LDGEC network with denoiser-by-denoiser training. SNR = $0$, $10$, and $15$ dB  are considered. From the figure, the LDGEC network with denoiser-by-denoiser training converges within $6$ layers, and more layers are required when the SNR is increasing.

{\blc
\emph{2) Without  real channel data}: Fig. \ref{fig:LDGEC_iterations}(b) demonstrates the convergence of the LDGEC-based channel estimator with layer-by-layer training. Specifically, the LDGEC-MSE means training the LDGEC network with MSE loss function  while LDGEC-SURE indicates training LDGEC network with SURE loss.  From Fig.\,\ref{fig:LDGEC_iterations}(b), the NMSE performance of LDGEC-SURE is close to that of LDGEC-MSE  when SNR = $10$ dB. On the contrary, the performance gap is approximately $2.5$ dB  when SNR = $0$ dB because the estimate of equivalent noise variance $v_{1h}$ and Monte-Carlo approximation of divergence (\ref{eqn:div_est}) are not accurate enough in the low-SNR regime, thereby degrading the channel estimation performance.
}
\subsection{Performance comparison}
\begin{figure}
\begin{minipage}{3in}
  \centerline{\includegraphics[width=3.0in]{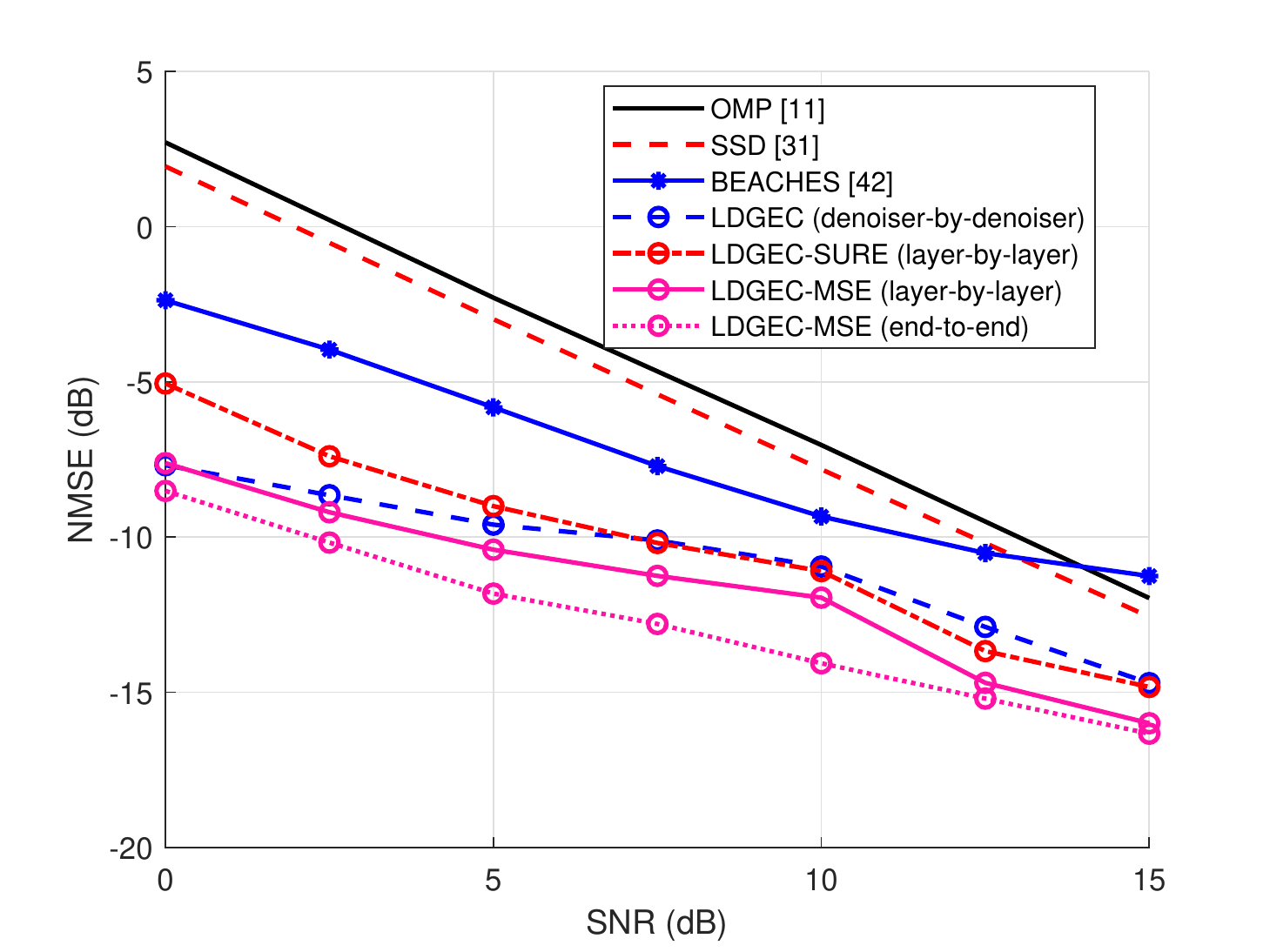}}
  \centerline{(a) $M=64$}\label{fig:channela}
\end{minipage}
\hfill
\begin{minipage}{3in}
  \centerline{\includegraphics[width=3.0in]{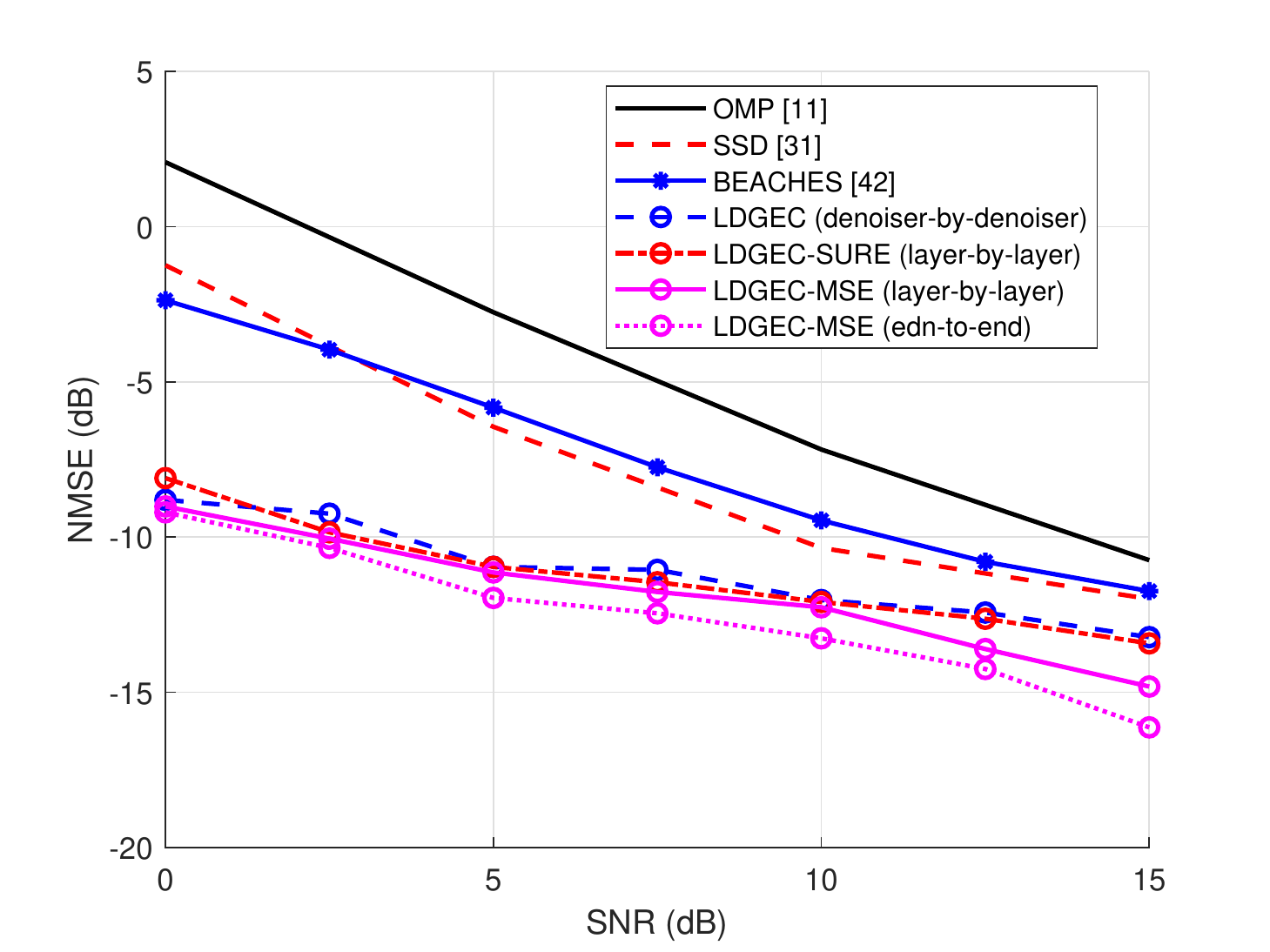}}
  \centerline{(b) $M=128$} \label{fig:SURE_sim}
\end{minipage}
\caption{.~~NMSEs performance comparisons of the LDGEC network with other channel estimation algorithms.}
\label{fig:LDEP_sim}
\end{figure}

Fig.\,\ref{fig:LDEP_sim} compares the performance of the LDGEC network  with other channel estimation algorithms. For LDGEC network with denoiser-by-denoiser, we set the number of layers $T=20$ for all SNR. For the LDGEC-SURE and LDGEC-MSE, we set the number of layers $T=20$ for SNR $\leq 10$ dB while $T=40$ for SNR $ > 10$ dB, because the LDGEC with layer-by-layer training needs more layers to converge in higher SNR. {\blc $M=64$ and $128$ are considered in the simulation.} From the figure, the LDGEC-based channel estimator can outperform the traditional CS-based algorithms with different training methods, such as OMP \cite{alkhateeb2014channel}, SSD \cite{beam2019Gao}, BEACHES \cite{BEACHES}. Note that the LDGEC  with layer-by-layer training can outperform that with denoiser-by-denoiser training {\blc if the MSE is considered as the loss function} because we need to divide equivalent noise variance $\hat{\sigma}^{2}$ into several intervals and train one DnCNN denoiser for each interval in denoiser-by-denoiser training, respectively. Instead of using the coarse intervals, the layer-by-layer training employs the accurate equivalent noise variance estimate, $v_{1h}$, in each layer, thereby improves the denoising performance.

\subsection{Impact of measurement ratio}

\begin{figure}
\begin{minipage}{3in}
  \centerline{\includegraphics[width=3.0in]{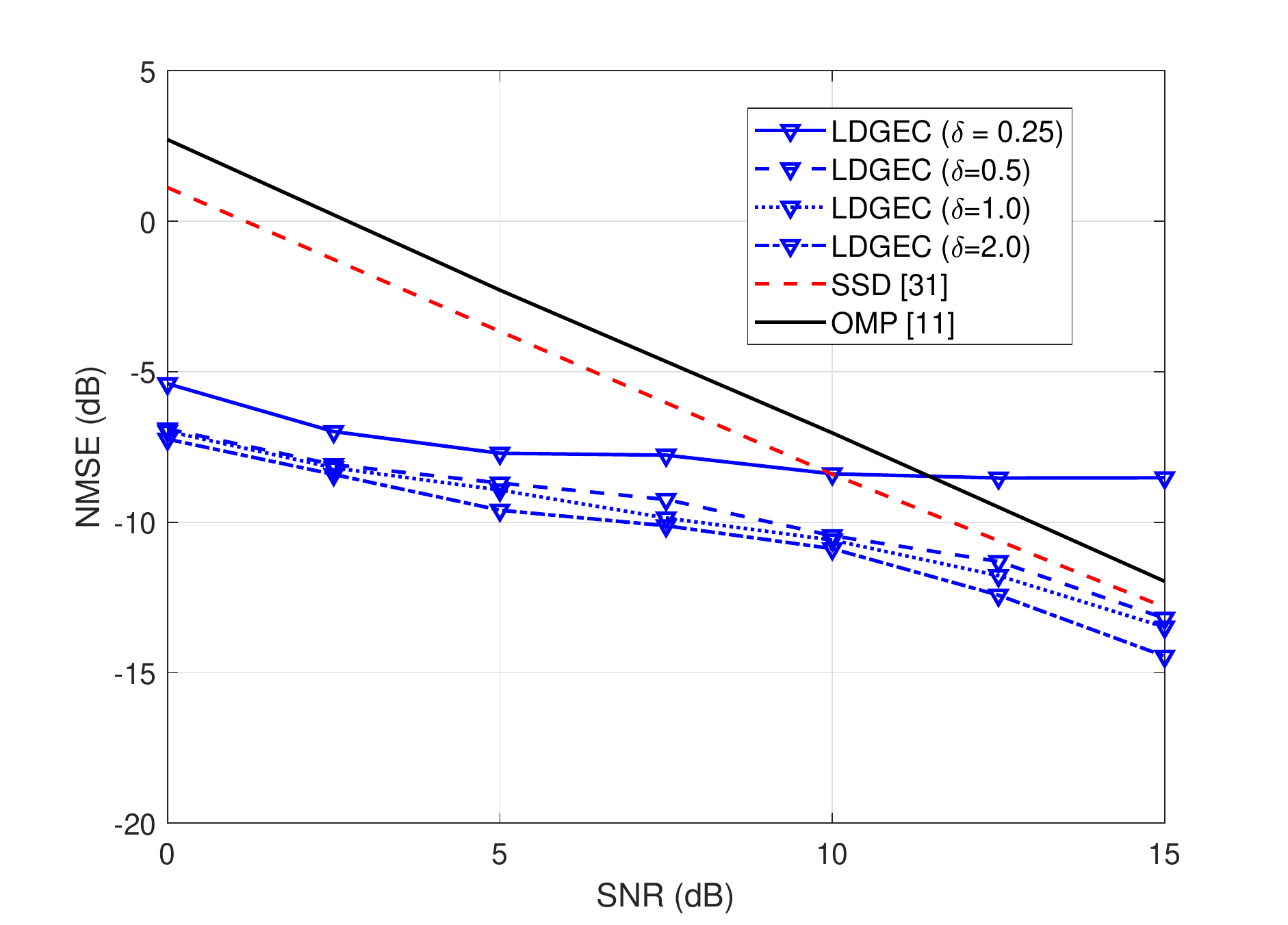}}
  \centerline{(a) Denoiser-by-denoiser training}\label{fig:channela}
\end{minipage}
\hfill
\begin{minipage}{3in}
  \centerline{\includegraphics[width=3.0in]{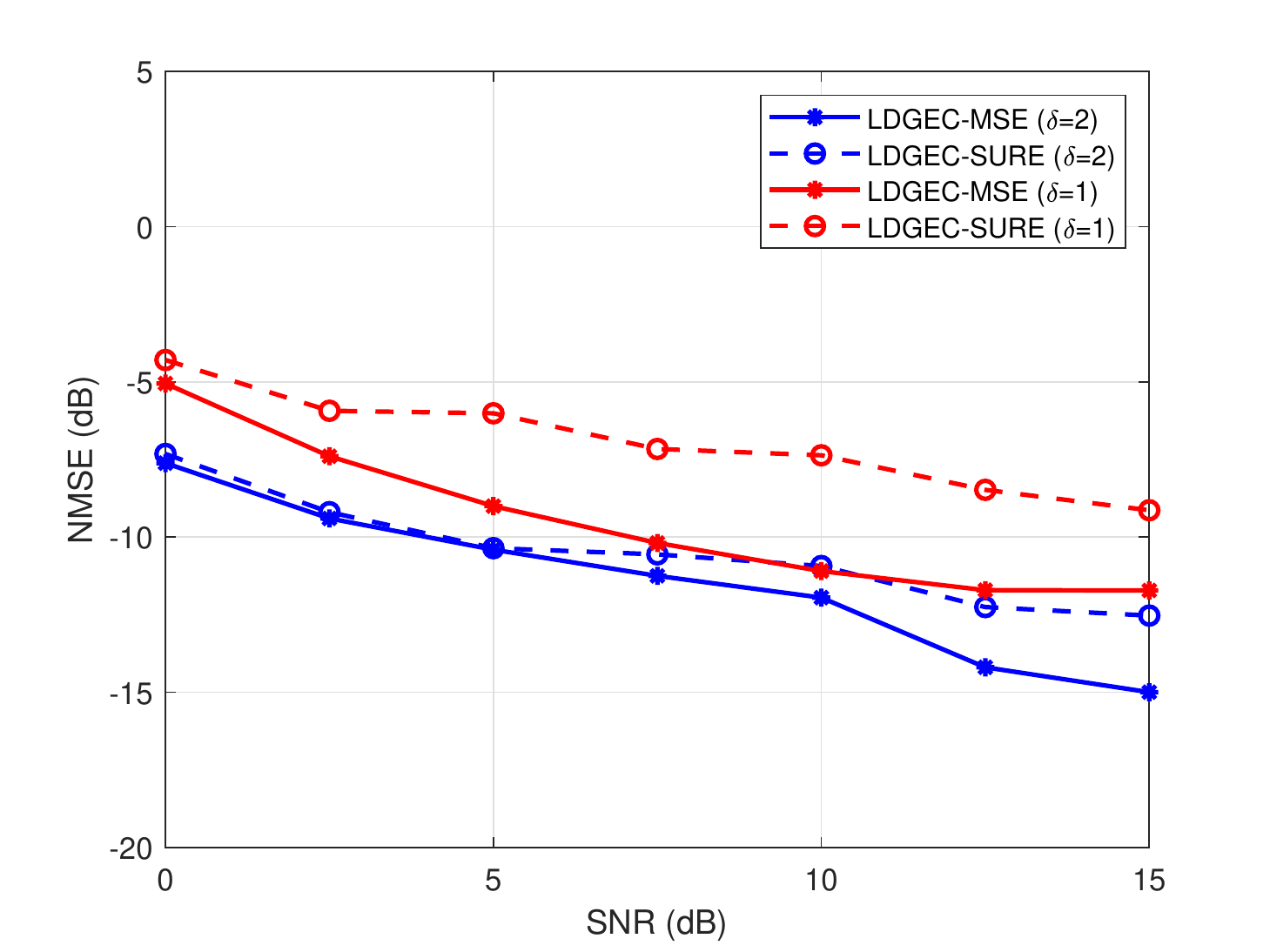}}
  \centerline{(b) Layer-by-layer training} \label{fig:SURE_sim}
\end{minipage}
\caption{.~~NMSE performance of LDGEC network with different measurement ratios for wideband beamspace mmWave MIMO systems.}
\label{fig:LDAMP_MR}
\end{figure}

In Section \ref{system_model}, the measurement ratio, defined by $\delta = Q N_{RF}/N$ and involved in the number of RF chains, $N_{RF}$, and pilot length $Q$, influences the performance of channel estimation and related to the system overhead. Fig.\,\ref{fig:LDAMP_MR}(a) illustrates the performance of the LDGEC network with denoiser-by-denoiser training versus different measurement ratios. {\blc Since SSD algorithm cannot work when $\delta \leq 1$, we consider  $\delta = 2$.} From the figure, the performance of the LDGEC network improves as the measurement ratio increases. Interestingly, the LDGEC algorithm with $\delta = 1$ outperforms the SSD algorithm with $\delta = 2$. Furthermore, the performance of the LDGEC network with $\delta = 1$ is close to that with $\delta = 2$, which demonstrates the strong robustness to the reduced number of RF chains. As the measurement ratio is determined by $N_{RF}$ and $Q$, we can decrease the number of RF chains $N_{RF}$ by increasing the number of pilot length $Q$, which can reduce hardware cost and power consumption of the system significantly. Fig.\,\ref{fig:LDAMP_MR}(b) illustrates the performance of the LDGEC network with layer-by-layer training versus different measurement ratios. From the figure, we have similar conclusions to that of LDGEC with denoiser-by-denoiser training.

\subsection{Low-resolution ADC}
The lens-based beamspace mmWave system can decrease the hardware cost by reducing the number of RF chains. However, a common limitation of the architectures is that the receiver RF chains include high-resolution ADCs, which are power-hungry devices, especially when large bandwidth is involved. The power consumption of a typical ADC roughly scales linearly with the bandwidth and grows exponentially with the quantization bits \cite{lee2008analog}. Many researchers have studied the mmWave massive MIMO systems with low-resolution ADCs\cite{Mo14ACSSP,JSTSP18He,Zhang2017mixed}. In this subsection, we investigate the LDGEC-based wideband beamspace channel estimation with low-resolution ADCs.

\begin{figure}[h]
\begin{minipage}{3in}
  \centerline{\includegraphics[width=3.0in]{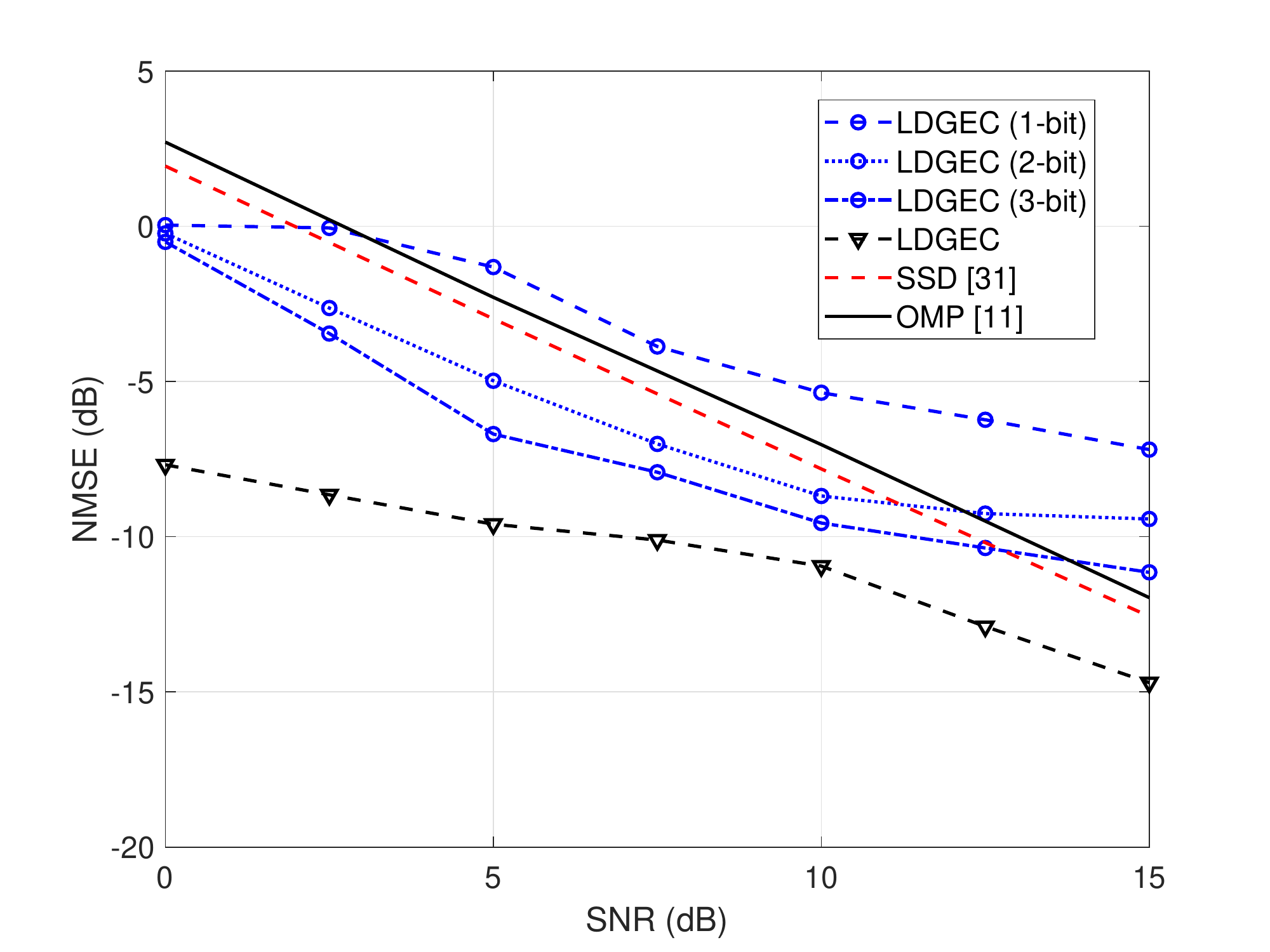}}
  \centerline{(a) Denoiser-by-denoiser training}\label{fig:LDGEC_sim1}
\end{minipage}
\hfill
\begin{minipage}{3in}
  \centerline{\includegraphics[width=3.0in]{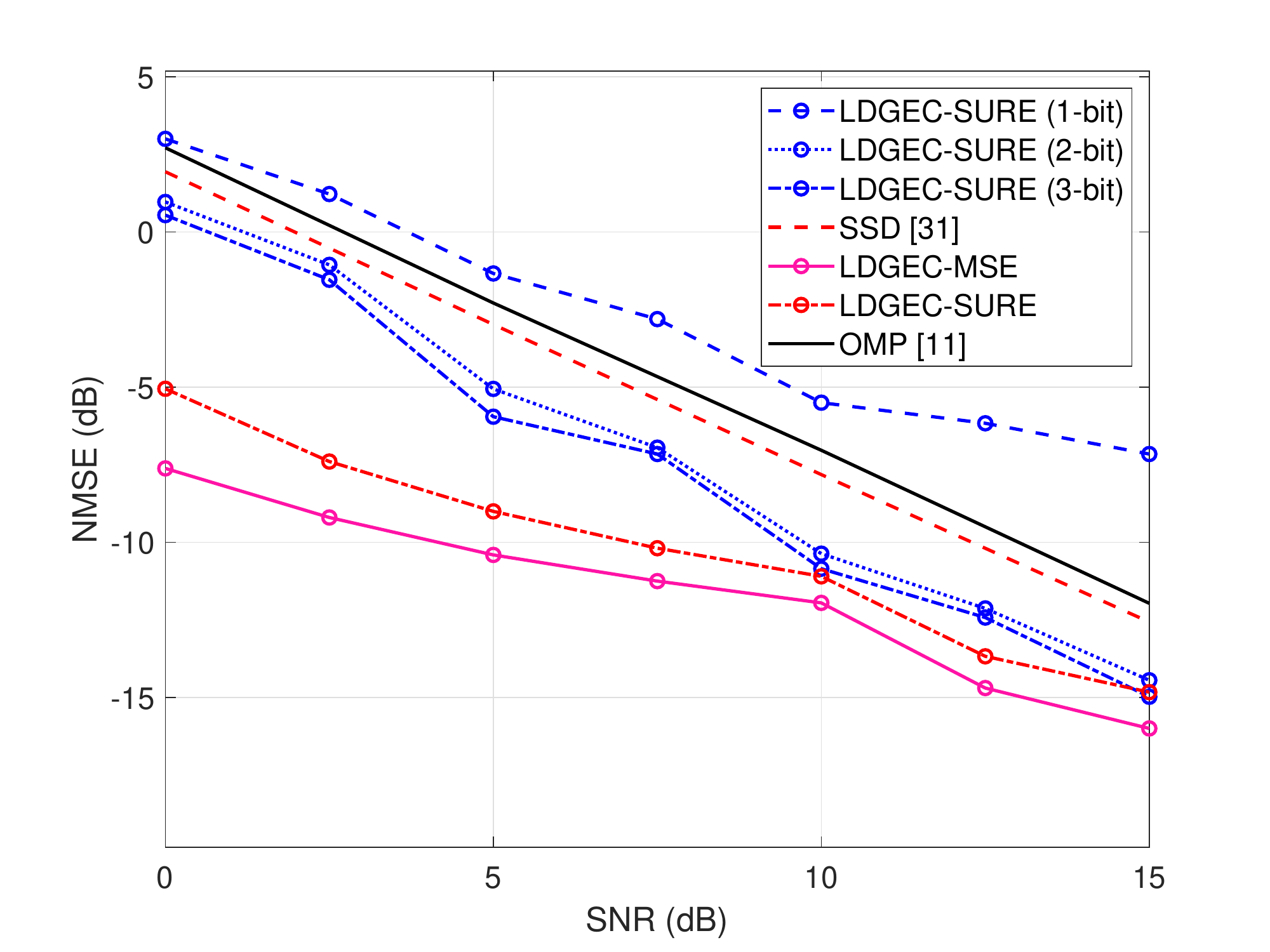}}
  \centerline{(b) Layer-by-layer training} \label{fig:LDGEC_sim2}
\end{minipage}
\caption{.~~NMSE performance of LDGEC network for wideband beamspace mmWave MIMO systems with low-resolution ADC.}
\label{fig:LDGEC_sim}
\end{figure}

To improve the robustness of LDGEC network in quantized systems, we use the damping method presented {\blc in} \cite{TSP20Wang} where damping factor $\beta=0.1^{t}$ is exponentially decreased. Fig.\,\ref{fig:LDGEC_sim}(a) compares the performance of the LDGEC channel estimator with low-resolution ADCs. From the figure, the LDGEC channel estimator with two-bit ADCs outperforms the SSD algorithm with infinite-bit ADCs when $\mathrm{SNR}<10$ dB. Therefore, the LDGEC channel estimator can accurately estimate the channel from the quantized signal, thereby reducing the hardware cost of the systems.

Fig.\,\ref{fig:LDGEC_sim}(b) illustrates the performance of the LDGEC network with SURE loss and layer-by-layer training. From the figure, we have similar conclusions to that of LDGEC with denoiser-by-denoiser training. Furthermore, the LDGEC-SURE with two-bit ADCs outperforms the SSD algorithm with infinite-bit ADCs. Therefore, the LDGEC-SURE is a promising approach to perform beamspace channel estimation in mmWave systems with low-resolution ADC architecture even without  real channel data.


\section{Conclusion}\label{con}

We have developed a novel model-driven unsupervised  DL network for wideband mmWave beamspace channel estimation. This network inherits the superiority of iterative signal recovery algorithms and the advanced DL-based denoiser, and thus presents excellent performance. The LDGEC network is easy to train and can be applied to a variety of selection networks. Furthermore, By utilizing the SURE loss, the LDGEC network can be trained without real channel data, enables the system to apply in a new environment. Simulation results demonstrate that the LDGEC-based channel estimator significantly outperforms state-of-the-art CS-based algorithms even for the receiver is equipped with a small number of RF chains and low-resolution ADCs. {\blc For future work, it will be interesting to apply the SURE technology to other DL-based wireless communication applications, such as CSI feedback and hybrid beamforming problems, to achieve unsupervised learning.}


\end{document}